\documentclass[acmsmall,screen]{acmart}
\usepackage{graphicx} 
\usepackage{appendix}
\usepackage[cache=false]{minted}
\usepackage{listings}
\usepackage{xcolor}
\usepackage{tikz}
\usepackage{caption}
\usepackage{subcaption}
\usepackage{pgfplots}
\pgfplotsset{compat=1.18}
\usetikzlibrary{calc, shapes.geometric, arrows, patterns}

\usetikzlibrary{shapes.geometric}

\tikzset{
    use bounding box relative coordinates/.style={
        shift={(current bounding box.south west)},
        x={(current bounding box.south east)},
        y={(current bounding box.north west)}
    },
}

\usepackage{algpseudocode}
\usepackage{algorithm}

\lstdefinestyle{myStyle}{
    captionpos=b,
    breaklines=true,
    frame=single,
    numbers=left,
    numberstyle=\tiny\color{red},
    numbersep=-2em,
    basicstyle=\scriptsize\ttfamily,
    keywordstyle=\bfseries\color{green!40!black},
    commentstyle=\itshape\color{purple!40!black},
    identifierstyle=\color{blue},
    backgroundcolor=\color{gray!10!white},
    belowskip=0pt,
}

\lstdefinelanguage{Datalog}{
  morekeywords=[1]{not,exists},
  morekeywords=[2]{decl, symbol, Operand, Instruction, LineNumber, output},
  morekeywords=[3]{choice, domain, delimiter, },
  morecomment=[l]{\%},
  sensitive=true
}
\lstdefinelanguage{plain}{
  morekeywords={},
  sensitive=false,
  morecomment=[l]{},
  morestring=[b]{}
}

\newenvironment{packed_itemize}{
\begin{list}{\labelitemi}{\leftmargin=2em}
 \setlength{\itemsep}{0pt}
 \setlength{\parskip}{0pt}
 \setlength{\parsep}{0pt}
}{\end{list}}

\algnewcommand\algorithmicinput{\textbf{Input:}}
\algnewcommand\INPUT{\item[\algorithmicinput]}
\algnewcommand\algorithmicoutput{\textbf{Output:}}
\algnewcommand\OUTPUT{\item[\algorithmicoutput]}
\usepackage{url}
\usepackage{xspace}
\newcommand{\tool}{\textsc{PAGENT}\xspace}
\title{Program Analysis Guided LLM Agent for Proof-of-Concept Generation}

\author{Achintya Desai }
\affiliation{\institution{University of California, Santa Barbara}  \city{Santa Barbara} 
\state{CA} \country{USA}}
\email{achintya@ucsb.edu}

\author{Md Shafiuzzaman}
\affiliation{\institution{University of California, Santa Barbara}  \city{Santa Barbara} 
\state{CA} \country{USA}}
\email{mdshafiuzzaman@ucsb.edu}

\author{Wenbo Guo}
\affiliation{\institution{University of California, Santa Barbara}  \city{Santa Barbara} 
\state{CA} \country{USA}}
\email{henrygwb@ucsb.edu}

\author{Tevfik Bultan}
\affiliation{\institution{University of California, Santa Barbara}  \city{Santa Barbara} 
\state{CA} \country{USA}}
\email{bultan@ucsb.edu}
\date{April 2026}

\begin{document}

\maketitle

Software developers frequently receive vulnerability reports that require them to reproduce the vulnerability in a reliable manner by generating a proof-of-concept (PoC) input that triggers it. 
Given the source code for a software project and a specific code location for a potential vulnerability, automatically generating a PoC for the given vulnerability has been a challenging research problem. 
Symbolic execution and fuzzing techniques require expert guidance and manual steps and face scalability challenges for PoC generation. 
Although recent advances in LLMs have increased the level of automation and scalability, the success rate of PoC generation with LLMs remains quite low.
In this paper, we present a novel approach called {\em Program Analysis Guided proof of concept generation agENT (PAGENT)} that is scalable and significantly improves the success rate of automated PoC generation compared to prior results.
PAGENT integrates lightweight and rule-based static analysis phases for providing static analysis guidance and sanitizer-based profiling and coverage information for providing dynamic analysis guidance with a PoC generation agent. Our experiments demonstrate that the resulting hybrid approach significantly outperforms the prior top-performing agentic approach by 132\% for the PoC generation task.

\section{Introduction}
Due to increasing interdependencies within software ecosystems it is crucial to detect and mitigate security vulnerabilities as quickly as possible to restrict their impact.
With the development of sophisticated vulnerability-detection techniques, the number of discovered vulnerabilities has also risen significantly in recent years~\cite{cvedata}.
When a vulnerability is discovered in a software project, it is reported to the project's developers and is expected to result in a patch that fixes it.
Given the rising number of vulnerabilities, addressing them in a timely manner can be challenging and critical, especially in open-source software projects.
Reproducing the vulnerabilities is an important step in addressing them through vulnerability triage, patch generation, and patch validation.

Reproducing a vulnerability in a reliable manner involves generating a proof-of-concept (PoC) input that triggers it.
The existence of a PoC input primarily proves the presence of a vulnerability.
It would be natural to assume that each reported vulnerability is supported by a PoC, which, surprisingly, is not the case.
Vulnerability reports lacking a valid PoC are not uncommon.
It often becomes the developer's responsibility to assess these vulnerability reports and produce a reliable PoC, which is a time-consuming task.
The key technical challenge in the PoC generation task, in general, is achieving an automated, scalable combination of vulnerability-detection capabilities to identify the vulnerable code snippet and semantic understanding of the codebase to infer the inputs that can reach the vulnerable code location.
Traditional static analysis tools are scalable but are widely known to be prone to false positives and do not automatically produce a PoC.
Although symbolic execution~\cite{baldoni2018survey} has the ability to produce a PoC and a precondition for the detected vulnerability, it suffers from the path explosion problem and often relies on manual environmental modeling and expert guidance to scale.
Traditional fuzzing is the most widely used automated vulnerability-detection technique that generates PoC inputs.
However, fuzzers also require expert guidance via fuzzing harness, input grammar, dictionaries, protocol awareness, and high-quality seed corpora to be effective.

Hybrid vulnerability-detection techniques have shown potential for minimizing the drawbacks of each technique while enhancing their strengths.
Fuzzing and static analysis have been shown to be complementary approaches for detecting memory-safety-related bugs~\cite{hassler2025comparative}.
There have been efforts to integrate these approaches in the literature~\cite{shastry2017static,wustholz2020targeted,saha2023rare,zheng2019efficient}. 
Static analysis has also been used to guide symbolic execution to achieve scalability and reduce false positives in vulnerability detection~\cite{aslanyan2024combining,shafiuzzaman2024stase}.
However, they all rely on a human-in-the-loop to provide expert knowledge to either guide the tool or convert the tool's findings into a vulnerability-triggering PoC input.
The application of Large Language Models (LLMs) to software engineering tasks~\cite{liu2023your,feng2024prompting} is becoming increasingly popular.
LLM agents are already becoming a viable direction for increasing automation and scalability of existing solutions.
Across various datasets, LLM agents~\cite{jain2024llm} have also demonstrated a potential ability to analyze code and infer its semantics.
However, as of now, LLM agents have shown poor accuracy due to their tendency to hallucinate.
This is a major challenge to the applicability of agents for PoC generation, as they are likely to produce inaccurate results without guidance.

Inspired by prior hybrid analysis approaches, and to address the challenges involved in automated PoC generation, we propose a novel hybrid approach that we call {\em Program Analysis Guided LLM agENT (PAGENT)} for PoC generation.
PAGENT takes a project's source code, a target code location, and a project build script as input and automatically generates a PoC if a vulnerability exists at the target code location, using static and dynamic analysis-guided LLM agents.
PAGENT is composed of three components: 1) Static Analysis, 2) PoC Generation Agent, and 3) Dynamic Analysis.
The primary goal of the static analysis stage is to produce reliable vulnerability-specific guidance for the PoC generation agent from the source code.
Towards this goal, we desire the following properties from our static analysis approach: 1) scalable towards large codebases, and 2) customizable and extensible to support vulnerability patterns.
The static analysis component generates the vulnerability report by extracting vulnerability-relevant information with source code-level program analysis.
The vulnerability entry from the input code location is then fed to the LLM agent, which has interactive access to the source code to generate a candidate PoC.
The candidate PoC is executed in a test environment that hosts an instrumented binary for dynamic analysis.
If the candidate PoC fails to trigger a crash, dynamic analysis provides feedback to the agent with respect to the execution run in the test environment.
The agent utilizes this feedback to further refine its code analysis to craft a valid PoC.
This loop between the agent and the dynamic analysis continues until either a PoC that triggers a crash is generated or the iteration budget is exhausted.

We evaluate PAGENT on a dataset of 203 vulnerabilities, chosen from Cybergym~\cite{wang2025cybergym}, spanning 10 open-source software projects across diverse software domains, including an IoT protocol, a Data compression library, and GNU binary tools.
Given the textual vulnerability report and the corresponding stack trace as input, the GPT-5 agent achieves the highest performance among the competing agents from the Cybergym benchmark.
PAGENT, without a stack trace or textual report, with the DeepSeek3.1-termius model, one of the weakest-performing models at this task, outperforms both Sonnet-4 and GPT-5 agents, achieving an overall accuracy of 42\% on our dataset.
PAGENT with DeepSeek3.2, a cheaper, open-source alternative to closed-source models, achieves 64.6\% accuracy on the same dataset and significantly outperforms the GPT-5 agent, our best results.
Furthermore, PAGENT identifies 32 post-patch vulnerabilities that trigger in the patched version of the source code as well.

In software codebases, it is common to maintain a shared repository for source code.
As the developers maintain and add new features to the software through code commits, PAGENT can leverage code locations from the change log to detect any vulnerabilities introduced through the corresponding commit.
The PAGENT approach can be easily integrated into CI/CD software pipelines for deployment after each commit.
For each code commit to the codebase, the modified code locations can be collected and directly fed into PAGENT, along with the source code.
If a code commit introduces a modeled vulnerability, PAGENT can detect and prove its presence by generating a concrete PoC input that triggers it.

Our overall contributions can be summarized as follows:
\begin{packed_itemize}
    \item \textbf{Static analysis for LLM agent:} We design and implement a scalable and automatable static analysis approach to guide LLM agents to improve the precision of LLM agents at PoC generation.
    \item \textbf{Dynamic analysis for LLM agent:} Our dynamic analysis phase drives the PoC validation, a crucial step in confirming the vulnerability, to facilitate the LLM agent to refine its code analysis.
    \item \textbf{Enhancing agents' effectiveness:} PAGENT improves the accuracy of the worst-performing open-sourced LLM (DeepSeek3.1) with static and dynamic program analysis by 132\%, and it outperforms the top 2 best-performing closed-sourced LLMs (Sonnet 4 and GPT-5-Reasoning) models by 100\%, and 104\% on average at a cheaper (32x, 0.42\$/vs 14\$/) cost per million output tokens.
    \item \textbf{Improvement at post-patch vulnerability detection:} PAGENT also demonstrates significant improvement (4x) at detecting post-patch vulnerabilities compared to the best-performing LLM agent on the dataset.
\end{packed_itemize}

\section{Overview}

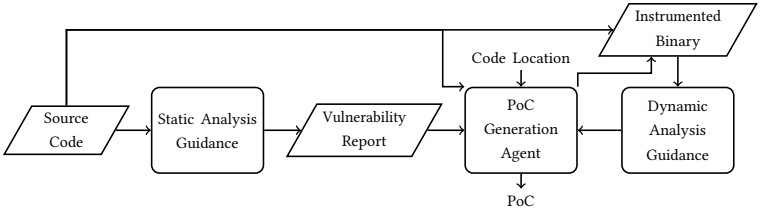
\begin{figure}[t!]
\centering
\scalebox{0.75}{
\begin{tikzpicture}[
rectnode/.style={rectangle, minimum height=30, draw=black, thick, minimum width=0.1*\textwidth, text width=3cm, align = center},
parallelonode/.style={trapezium, trapezium left angle=60, trapezium right angle=120, draw=black, thick, minimum height=1, minimum width=1cm, text width=1cm, align = center},
diamondnode/.style={diamond, draw=black, thick, minimum height=1, minimum width=0.1*\textwidth, text width=3cm, align = center},
roundedrectnode/.style={rounded corners, draw=black, thick, minimum height=1, minimum width=0.1*\textwidth, text width=3cm, align = center}
]
\useasboundingbox (0,0) rectangle (\textwidth,4);
\begin{scope}[use bounding box relative coordinates]
\node[parallelonode](sourcecode) at (0.1,0.36)[text width=1cm] {\footnotesize Source Code};
\node[roundedrectnode](staticguide) at (0.28,0.36)[text width=1.75cm, minimum height = 1.5 cm] {\footnotesize Static Analysis Guidance};
\node[parallelonode](vulreport) at (0.48,0.36)[text width=1.45cm] {\footnotesize Vulnerability Report};
\node[roundedrectnode](pocagent) at (0.68,0.36)[text width=1.75cm, minimum height = 1.5 cm] {\footnotesize PoC Generation Agent};
\node(codeloc) at (0.68,0.68) [text width=3cm, align = center] {\footnotesize Code Location};
\node[roundedrectnode](dynamicguide) at (0.88,0.36)[text width=1.75cm, minimum height = 1.5 cm] {\footnotesize Dynamic Analysis Guidance};
\node[parallelonode](instrubin) at (0.88,0.8)[text width=1.5cm] {\footnotesize Instrumented Binary};
\node(poc) at (0.68,0.05) [text width=3cm, align = center] {\footnotesize PoC};
\draw[->, thick] (sourcecode.east) -- (staticguide.west);
\draw[->, thick] (staticguide.east) -- (vulreport.west);
\draw[->, thick] (vulreport.east) -- (pocagent.west);
\draw[->, thick] (dynamicguide.west) -- (pocagent.east);
\draw[->, thick] (codeloc.south) -- (pocagent.north);
\draw[->, thick] (sourcecode.north) |- (instrubin.west);
\draw[->, thick] (pocagent.south) -- (poc.north);
\draw[->, thick] (pocagent.north east) |- (0.78,0.61) -| (instrubin.south west);
\draw[->, thick] (instrubin.south) -- (dynamicguide.north);
\draw[->, thick] (sourcecode.north) |- (0.58,0.8) |- (pocagent.north west);
\end{scope}
\end{tikzpicture}
}
\caption{Overview of~\tool Technique}
\label{fig:GuidedPoCAgentOverview}
\end{figure}

\begin{figure}[t!]
\scalebox{0.75}{
    \begin{minipage}{0.48\linewidth}
    \begin{tikzpicture}
      \def\cardwidth{\linewidth}
      \def\cardheight{5.5cm}
      \def\titleheight{0.5cm}
    
      \draw[rounded corners, black, line width=1pt]
        (0,0) rectangle (\cardwidth,\cardheight);
    
    
      \fill[black]
      (0,\cardheight-\titleheight) --
      ++(\cardwidth,0)
        {[rounded corners=4pt] --
         ++(0,\titleheight) --
         ++(-\cardwidth,0)}
      -- cycle;

      \node[anchor=west, text=white, align=left]
        at (0,\cardheight-\titleheight/2)
        {Static Analysis Guidance};
    
      \node[
        anchor=north west,
        text=black,
        align=left,
        text width=\cardwidth-0.75em,
      ] at (0,\cardheight-\titleheight)
      {\tt \footnotesize `Vulnerability Type': `Stack-Buffer-Overflow-Vulnerability',\\
        `Vulnerable Function': {\tt \footnotesize `get\_register\_operand'},\\
        `Entrypoint': {\tt \footnotesize `LLVMFuzzerTestOneInput'},\\
        `Taint Path': "[{\tt \footnotesize`LLVMFuzzerTestOneInput', \textbf{`print\_insn\_tic30'}, `print\_branch.171581',
        `get\_register\_operand'}]",\\
        `Vulnerable Program Location': `204',\\
        `Template Assertion Violation': {\tt \footnotesize `0 $<=$ get\_register\_operand:30:0:0 $<=$ SIZEOF(get\_register\_operand:\%25)'}
      };
    
    \end{tikzpicture}
    \end{minipage}
    \begin{minipage}{0.5\linewidth}
    \begin{tikzpicture}
      \def\cardwidth{\linewidth}
      \def\cardheight{5.5cm}
      \def\titleheight{0.5cm}
    
      \draw[rounded corners, black, line width=1pt]
        (0,0) rectangle (\cardwidth,\cardheight);
    
    
      \fill[black]
      (0,\cardheight-\titleheight) --
      ++(\cardwidth,0)
        {[rounded corners=4pt] --
         ++(0,\titleheight) --
         ++(-\cardwidth,0)}
      -- cycle;

      \node[anchor=west, text=white, align=left]
        at (0,\cardheight-\titleheight/2)
        {Dynamic Analysis Guidance};
    
      \node[
        anchor=north west,
        text=black,
        align=left,
        text width=\cardwidth-0.75em,
      ] at (0,\cardheight-\titleheight)
      {\tt \footnotesize
        [...
        {`file\_path":"/src/binutils-gdb/opcodes/
        tic4x-dis.c","function\_name":\textbf{"print\_insn\_tic4x"},
        `region\_coverage":75.00,"line\_coverage":80.95,
        "branch\_coverage":11.76},
        {`file\_path":"/src/binutils-gdb/opcodes/
        tic4x-dis.c","function\_name":`tic4x-dis.c:tic4x\_
        disassemble", "region\_coverage":64.71,"line\_
        coverage":77.05,"branch\_coverage":16.67},
        {`file\_path":"/src/binutils-gdb/opcodes/
        tic4x-dis.c","function\_name":"tic4x-dis.c:
        tic4x\_print\_register","region\_coverage":72.41,
        "line\_coverage":83.33,"branch\_coverage":16.67},\\
        ...]
      };
    
    \end{tikzpicture}    
    \end{minipage}
    }
    \caption{Static and Dynamic analysis guidance example for ARVO:18615}
    \label{fig:motivation}
    
\end{figure}
We aim to address the following problem: \textit{Given a source code $\mathcal{S}$ and a specific code location $L$,  automatically generate a functional Proof-of-Concept (PoC) input $I$ that demonstrates the existence of a vulnerability at location $L$.} To address this problem, we propose a hybrid framework, illustrated in Fig.~\ref{fig:GuidedPoCAgentOverview}, that integrates static and dynamic analysis with an LLM-based agent. The framework starts with a two-phase static analysis that generates an entrypoint-driven reachability graph using a light-weight static analysis phase, followed by a scalable, customizable, and extensible rule-based static analysis phase.  An LLM-driven PoC generation agent then leverages these results to navigate the codebase to resolve the input constraints required to synthesize a candidate PoC. This candidate is dispatched via a bash command to a test environment where it is executed against a sanitizer-instrumented binary  of $\mathcal{S}$. In the event of a failure, the agent receives the dynamic analysis results in a command response and a coverage report file. The agent utilizes this feedback to iteratively refine the input until a successful PoC is produced or the iteration budget is exhausted.



Fig.~\ref{fig:motivation} showcases an example of static and dynamic analysis guidance produced by our framework for ARVO:18615 vulnerability instance from the dataset. The instance focuses on a buffer overread vulnerability in the \texttt{get\_register\_operand} function of the GNU Binutils TIC30 disassembler. As shown in Listing~\ref{lst:motivating-example}, the function invokes \texttt{strncpy} with a fixed copy length \texttt{OPERAND\_BUFFER\_LEN}, although the destination buffer allocated by its caller may be smaller. This size mismatch results in a buffer overread. Triggering this vulnerability requires not only identifying the faulty copy operation but also constructing an input $I$ that selects the correct disassembly mode and reaches the vulnerable execution context. Automatically synthesizing such an $I$  requires reasoning over complex, cross-file program logic, and input-dependent control flow. This task is well-suited for an LLM-based PoC agent, but difficult to encode with fixed heuristics. Relying solely on natural-language vulnerability descriptions is insufficient in such cases. For example, the vulnerability report for this instance says “An array overrun occurs in \texttt{tic30-dis.c} within the \texttt{print\_branch} function due to an incorrect size of the operand array when disassembling corrupt TIC30 binaries”. This guides the agent to the relevant source file and vulnerable function,  but lacks the precise semantics including feasible entry points and call-path constraints needed to reach $L$. This limitation motivates the use of static analysis to extract vulnerability-specific guidance that grounds the agent’s reasoning in the program structure. Static analysis can explicitly encode feasible entrypoints and reachability constraints extracted from the program as illustrated in Fig.~\ref{fig:motivation}. This information enables the PoC generation agent to leverage call-path information and vulnerability metadata to select candidate instruction patterns and architecture constants intended to invoke \texttt{get\_register\_operand}.

\begin{lstlisting}[caption=Code snippet from tic30-dis.c file in binutils project, label={lst:motivating-example}, language=C,  firstnumber=193, escapechar=|]
    static int get_register_operand (unsigned char fragment, char *buffer)
    {
      const reg *current_reg = tic30_regtab;
      if (buffer == NULL) return 0;
      for (; current_reg < tic30_regtab_end; current_reg++) {
          if ((fragment & 0x1F) == current_reg->opcode){ 
              strncpy (buffer, current_reg->name, OPERAND_BUFFER_LEN);  |\label{line:tic30-vuln}| /* Vulnerable copy */
              buffer[OPERAND_BUFFER_LEN - 1] = 0;
              return 1;
            }  }
      return 0;
    }
\end{lstlisting}

However, static guidance alone can be imprecise when multiple execution paths coexist. In this example, the generated inputs may exercise the TIC4X disassembler (\texttt{tic4x-dis.c}) instead of the intended TIC30 disassembler (\texttt{tic30-dis.c}), due to an incorrect architecture selection. Dynamic analysis feedback resolves this ambiguity. By inspecting coverage information illustrated in Fig.~\ref{fig:motivation} from failed executions, the agent detects that the vulnerable code path is not being reached and identifies the incorrect architecture configuration. The agent then revises its input accordingly, selecting the correct TIC30 architecture constant. This refinement enables the generated PoC to reach \texttt{get\_register\_operand} and successfully trigger the buffer overread.

\section{Static and Dynamic Analysis Guided Agentic PoC Generation}
We describe the technical details of the three main components of PAGENT (Fig.~\ref{fig:GuidedPoCAgentOverview}) in this section.
\subsection{Static Analysis Guidance}

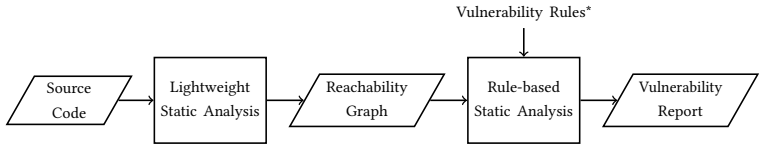
\begin{figure}[t!]
\centering
\scalebox{0.75}{
\begin{tikzpicture}[
rectnode/.style={rectangle, minimum height=30, draw=black, thick, minimum width=0.1*\textwidth, text width=3cm, align = center},
parallelonode/.style={trapezium, trapezium left angle=60, trapezium right angle=120, draw=black, thick, minimum height=1, minimum width=1cm, text width=1cm, align = center},
diamondnode/.style={diamond, draw=black, thick, minimum height=1, minimum width=0.1*\textwidth, text width=3cm, align = center},
roundedrectnode/.style={rounded corners, draw=black, thick, minimum height=1, minimum width=0.1*\textwidth, text width=3cm, align = center}
]
\useasboundingbox (0,0) rectangle (\textwidth,4);
\begin{scope}[use bounding box relative coordinates]
\node[parallelonode](sourcecode) at (0.1,0.5)[text width=1cm] {\footnotesize Source Code};
\node[rectnode](lightstat) at (0.28,0.5)[text width=1.75cm, minimum height = 1.5 cm] {\footnotesize Lightweight Static Analysis};
\node(rules) at (0.68,0.88) [text width=3cm, align = center] {\footnotesize Vulnerability Rules*};
\node[parallelonode](reachgraph) at (0.48,0.5)[text width=1.45cm] {\footnotesize Reachability Graph};
\node[rectnode](rulestat) at (0.68,0.5)[text width=1.75cm, minimum height = 1.5 cm] {\footnotesize Rule-based Static Analysis};
\node[parallelonode](vulnreport) at (0.88,0.5)[text width=1.45cm] {\footnotesize Vulnerability Report};
\draw[->, thick] (sourcecode.east) -- (lightstat.west);
\draw[->, thick] (lightstat.east) -- (reachgraph.west);
\draw[->, thick] (reachgraph.east) -- (rulestat.west);
\draw[->, thick] (rules.south) -- (rulestat.north);
\draw[->, thick] (rulestat.east) -- (vulnreport.west);
\end{scope}
\end{tikzpicture}
}
 \vspace*{-0.2in}
\caption{Overview of Static Analysis Guidance Component {\footnotesize(* denotes reusable input across software projects)}}
\label{fig:STAG}
\end{figure}

\begin{algorithm}[t!]
\footnotesize
\caption{Static Analysis Guidance Component }\label{alg:StatAnalAlgo}
\begin{algorithmic}[1]
\INPUT Source code $S$, Build Instructions $b$, Rule Repository $R$, (Optional) Entrypoints $E_p$
\OUTPUT Vulnerability Report $V_R$
\State \texttt{\hspace*{0.1in} // Phase 0: Compile the source code to LLVM-IR}
\State $S_L \leftarrow \texttt{Build($S$, $b$)}$
\State \texttt{\hspace*{0.1in} // Phase 1: Lightweight Static Analysis}
\State $E_p = E_p \cup \{\texttt{LLVMFuzzerTestOneInput()}, \texttt{main()}\}$
\State $E_{l} \leftarrow \texttt{Set\_Entrypoints($E_p$)}$
\State $G \leftarrow \texttt{Construct\_CallGraph($S_{L}$)}$
\State $T \leftarrow \texttt{Filter\_Reachable($G$, $E_{l}$)}$
\State \texttt{\hspace*{0.1in} // Phase 2: Rule-based Static Analysis}
\State $F \leftarrow \texttt{Generate\_Program\_Facts($S_{L})$}$
\State $V \leftarrow \texttt{Apply\_Vulnerability\_Rules($F$,$R$)}$
\State $V_R = \phi$
\For{\texttt{each $v$ in $V$}}
\State $e = \phi$
\State \texttt{$e$["Vulnerability Type"] $= v$["Vulnerability Type"]}
\State \texttt{$e$["Vulnerable Function"] $= v$["Vulnerable Function"]}
\State \texttt{$e$["Taint Path"] $=$ Extract\_Path[$T$, $e$["Vulnerable Function"]]}
\State \texttt{$e$["Entrypoint"] $=$ $e$["Taint Path"][0]}
\State \texttt{$e$["Vulnerable Program Location"] $= v$["Vulnerable Program Location"]}
\State \texttt{$e$["Template Assertion Violation"] $= v$["Template Assertion Violation"]}
\State $V_R = V_R \cup e$
\EndFor
\end{algorithmic}
\end{algorithm}

We designed our static analysis to achieve scalability towards the large codebases and customizability and extensibility to support a variety of vulnerability patterns.
To improve the scalability of the analysis, we perform an initial lightweight static analysis on the source code. This includes entrypoint-driven automated call-graph construction, over-approximate indirect call resolution, and dead-code removal.
To make our static analysis customizable and extensible, we implement a rule-based static analysis technique that identifies vulnerable code patterns in the sourcecode. This technique allows us to add and customize rules depending on the vulnerability patterns. It can be reused across domains and can be extended by adding new vulnerability patterns when new types of vulnerabilities are discovered. 
Furthermore, the rules are written at LLVM-IR level, allowing the user to capture all source-code level variants for a specific bug pattern.
Figure~\ref{fig:STAG} provides an overview of our static analysis component.

\subsubsection{Lightweight Static Analysis}
Software codebases are built on a large number of functions.
To achieve scalability in PoC generation, it is essential to filter out functions that do not influence the given vulnerability.
Identifying such functions accurately through static analysis is challenging.
Removal of a function that holds a vulnerability or impacts the exploitability of a vulnerability can lead to false negatives.
To address this issue, we propose a lightweight static analysis component that takes source code with entrypoints as input and generates a reachability graph containing functions reachable only from the entrypoints.
This involves constructing an over-approximate call graph and filtering out unreachable functions from the entrypoints.
Lightweight static analysis component ensures that we only report the vulnerabilities that are potentially reachable from the specified entrypoints.

We begin by setting the entrypoints in the input source code. Lightweight static analysis automatically detects the common entrypoints, specifically the ones used by fuzzers such as {\tt \footnotesize main()}, {\tt \footnotesize LLVMFuzzerTestOneInput()}. It also supports an optional user-defined list of entrypoints.
Once the entrypoints are set, we perform an entrypoint-driven call graph construction pass where all the functions that are reachable from the entrypoints are identified. 
In complex codebases, indirect function calls via function pointers or virtual method tables (C++) are typical for supporting callback or event handler functionality.
Entrypoint-driven call graph construction pass involves resolving direct and indirect function calls via function signature analysis (FSA).
In the literature, FSA~\cite{li2025redefining} has been widely regarded as a sound approach when type information is available.
Compared to modern approaches like Multi-Layer Type Analysis (MLTA)~\cite{lu2019does} and Type-based dependence analysis~\cite{lu2023practical}, FSA is highly scalable due to linear time signature matching between indirect call instructions and function signatures.
However, FSA approach is prone to produce a high number of false positives due to over-approximation.
For lightweight static analysis, the FSA approach provides a scalable and sound strategy for constructing an over-approximate call graph. 
Once the list of reachable functions is generated from the entrypoint-driven call graph construction pass, we perform code elimination by marking unreachable functions from the entrypoints.
This pass ensures that the resulting source code only consists of potentially reachable functions from the entrypoints and essential code definitions. 

\begin{lstlisting}[caption=Code snippet from objdump.c file in binutils project, label={lst:objdump-example}, language=C,  firstnumber=4285, escapechar=|]
    static void dump_bfd_private_header (bfd *abfd)
    {
      if (!bfd_print_private_bfd_data (abfd, stdout))
        non_fatal (_("warning: private headers incomplete: %s"),
    	       bfd_errmsg (bfd_get_error ()));
    }
\end{lstlisting}

We implemented our lightweight static analysis component at the LLVM-IR level.
For tailored program analysis, LLVM provides a comprehensive infrastructure for writing analysis passes.
It also allows us to maintain compatibility with our rule-based static analysis, which utilizes the inherent modularity of LLVM-IR to design customizable vulnerability rules.
This adds an extra step at the beginning of the lightweight static analysis to use the build instructions to compile the project into LLVM-IR.
Consider the code snippets related to motivating example shown in Listing~\ref{lst:objdump-example} and Listing~\ref{lst:vmsalpha-example}.
{\tt \footnotesize dump\_bfd\_private\_header()} is an internal function, encountered along the path towards vulnerability from motivating example, that calls {\tt \footnotesize bfd\_print\_private\_bfd\_data()} to display internal data specific to the object file's format from a binary file descriptor (BFD).
Depending on the object file type, {\tt \footnotesize bfd\_print\_private\_bfd\_data()} gets resolved to a target-specific function defined in the BFD target vector.
For OpenVMS Alpha, this call gets resolved to {\tt \footnotesize vms\_bfd\_print\_private\_bfd\_data()} if defined in the BFD target vector.
For this example, the lightweight static analysis component translates the source code to LLVM-IR.
It detects the indirect call instruction, during the first pass at the LLVM level, associated with the function call on line number 4287 from Listing~\ref{lst:objdump-example}.
It matches the normalized function signature for the indirect call: {\tt \footnotesize i1 (\%struct.bfd* , i8*)} with the function definition of {\tt \footnotesize vms\_bfd\_print\_private\_bfd\_data(): i1 @vms\_bfd\_print\_private\_bfd\_data(\%struct.bfd\* \%0, i8* \%1)} and adds it as a call edge from function {\tt \footnotesize bfd\_print\_private\_bfd\_data()} to {\tt \footnotesize vms\_bfd\_print\_private\_bfd\_data()}.
This ensures that the function remains {\tt \footnotesize vms\_bfd\_print\_private\_bfd\_data()} reachable from the entrypoint due to the possibility of an indirect call from {\tt \footnotesize dump\_bfd\_private\_header()} at runtime.
\subsubsection{Rule-based Static Analysis}
It is well-known that LLM agents are prone to hallucination~\cite{zhang2025mirage,lin2025llm}, which can cause them to report false positive vulnerabilities and incorrect PoCs.
Specifically, LLM agents suffer from poor accuracy at the task of extracting vulnerability-specific information from the codebase, which is crucial for crafting accurate PoC input as demonstrated by the motivating example.
To address this issue, we introduce a rule-based static analysis component that can reliably generate vulnerability-specific information such as taint path, taint source, vulnerable function, and assertion template, capturing the vulnerability as an assertion violation.
Rule-based static analysis component is driven by the vulnerability rule repository.
The vulnerability rule repository consists of vulnerability rules that target commonly known vulnerabilities, such as integer overflow and buffer overflow.
There are two main advantages to maintaining the vulnerability rule repository: 1) Vulnerability rules are reusable across domains; Once they are added to the rule repository, they will be added to any future static analysis runs, 2) The vulnerability rule repository can be easily extended to support any new common or domain-specific vulnerabilities.
We modeled the following 12 vulnerabilities as vulnerability rules within the rule repository: Heap-buffer-overflow, Stack-buffer-overflow, Global-buffer-overflow, Heap-buffer-underflow, Stack-buffer-underflow, Global-buffer-underflow, Division-by-zero, Integer-Overflow, Integer-Underflow, Out-of-bounds, Use-after-free, Double-free.

\begin{lstlisting}[caption=Code snippet from vms-alpha.c file in binutils project, label={lst:vmsalpha-example}, language=C,  firstnumber=8307]
    static bool vms_bfd_print_private_bfd_data (bfd *abfd, void *ptr)
    {
      FILE *file = (FILE *)ptr;
      if (bfd_get_file_flags (abfd) & (EXEC_P | DYNAMIC))  evax_bfd_print_image (abfd, file);
      else {
          if (bfd_seek (abfd, 0, SEEK_SET)) return false;
          evax_bfd_print_eobj (abfd, file);
        }
      return true;
    }
\end{lstlisting}

Rule-based static analysis approach utilizes Datalog to specify vulnerability patterns.
For a given program, these analysis rules instantiate fixpoint computations corresponding to evaluations of Datalog logic programs, resulting in the generation of facts (such as vulnerable code locations in the code). 
Although the underlying facts vary across different programs, rule-based static analysis approach crucially allows the vulnerability rules themselves to be reused across programs.
Our rule-based static analysis component generates facts over the LLVM-IR version of the source code, applying vulnerability rules to detect vulnerable code locations, and extracting vulnerability-specific information from the code.
We implemented our rule-based static analysis using the open-source cclyzerpp tool built on Soufflé, an expressive dialect of Datalog.
For generating facts, cclyzerpp provides built-in LLVM passes that populate fact relations with input program facts based on the abstract syntax tree (AST) of LLVM modules.
We utilize Soufflé's highly parallel engine to detect vulnerabilities and extract their respective assertion template, code locations, vulnerable functions, and taint sources.

\begin{lstlisting}[caption=Vulnerability rule for Out-of-Bounds Vulnerability, label={lst:oobrule-example}, language=Datalog, numbers=none]
.decl out_of_bounds_primitive(?type: symbol, ?assertion: symbol, ?func: Function, ?op1: Operand, ?op2: Operand, ?instr: Instruction, ?line:LineNumber) choice-domain (?func, ?line)
.output out_of_bounds_primitive(delimiter=",")

out_of_bounds_primitive(?type, ?assertion, ?func, ?op1, ?op2, ?instr, ?line) :-
    ?type = "Out-of-Bounds-Vulnerability",
    ?assertion = cat("0 <= ", to_string(?op2), "<=SIZEOF(", to_string(?op1), ")"),
    instr_func(?instr, ?func),
    indexaccessinstructions(?op1, ?op2, ?instr),
    instr_pos(?instr, ?line, ?col).
    
\end{lstlisting}

Listing~\ref{lst:oobrule-example} shows the vulnerability rule for the out-of-bounds vulnerability, discussed in the motivating example, at the instruction level.
The vulnerability rule is composed of three clauses: 1) The first clause sets the value of {\tt \footnotesize ?type} to vulnerability type, 2) The second clause sets the value of {\tt \footnotesize ?assertion} which holds the assertion template specific to the vulnerability type, 3) The third set of clauses involve detecting the vulnerability in the form of program facts and constraints.
In the above example, we set the {\tt \footnotesize ?type} to the constant string  "Out-of-Bounds-Vulnerability" to indicate the vulnerability type. The {\tt \footnotesize ?assertion} is set to a vulnerability-specific template that is instantiated based on the operands involved in the vulnerability. For the motivating example, the {\tt \footnotesize ?assertion} will be set to string {\tt \footnotesize 0 <= <\/tmp\/full\_project.bc>:evax\_bfd\_print\_dst:126:0:5 <= SIZEOF(<\/tmp\/full\_project.bc>:evax\_bfd\_print\_dst:\%106)}.
Notice that the assertion template involves values directly from LLVM-IR and is not replaced with their source code name.
Based on the availability of debug information and optimization level, it is not always possible to resolve the LLVM-IR operand names to their source code names.
Although we do not generate source-code level assertions, the assertion template generated above provides a vulnerability-specific condition to be violated by the prospective PoC.
As LLMs are well-versed with the LLVM syntax, they can reason about assertion templates and extract useful information, such as 5 in {\tt \footnotesize <\/tmp\/full\_project.bc>:evax\_bfd\_print\_dst:126:0:5} indicates the index 5 access from the buffer {\tt \footnotesize buf}.
Listing~\ref{lst:vulnreport-example} shows an example vulnerability report for an out-of-bounds vulnerability generated by the rule-based static analysis component. 
For each potential vulnerability, our vulnerability report includes the following information: Vulnerability type, Vulnerable function, Entrypoint, Taint path, Vulnerable program location, and Template assertion.
\begin{lstlisting}[caption=Vulnerability report for Out-of-Bounds Vulnerability from motivating example 1, label={lst:vulnreport-example}, language=XML, numbers=none]
"potential_target_<num>": {
    "Vulnerability Type": "Out-of-Bounds-Vulnerability",
    "Vulnerable Function": "evax_bfd_print_dst",
    "Entrypoint": "LLVMFuzzerTestOneInput.70743",
    "Taint Path": "['LLVMFuzzerTestOneInput.70743', 'bfd_close.136854','_bfd_archive_close_and_cleanup.9552', 'dump_bfd_private_header', 'vms_bfd_print_private_bfd_data', 'evax_bfd_print_image', 'evax_bfd_print_dst']",
    "Vulnerable Program Location": "7260",
    "Template Assertion Violation": "0 <= </tmp/full_project.bc>:evax_bfd_print_dst:126:0:5 <= SIZEOF(</tmp/full_project.bc>:evax_bfd_print_dst:%106)" }
    
\end{lstlisting}

\subsection{PoC Generation Agent}
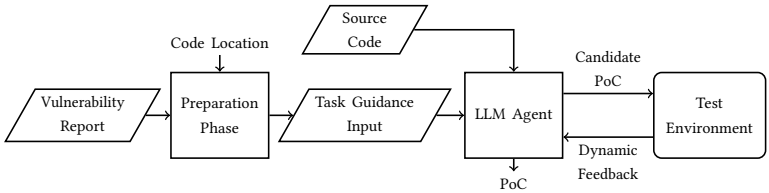
\begin{figure}[t!]
\centering
\scalebox{0.75}{
\begin{tikzpicture}[
rectnode/.style={rectangle, minimum height=30, draw=black, thick, minimum width=0.1*\textwidth, text width=3cm, align = center},
parallelonode/.style={trapezium, trapezium left angle=60, trapezium right angle=120, draw=black, thick, minimum height=1, minimum width=1cm, text width=1cm, align = center},
diamondnode/.style={diamond, draw=black, thick, minimum height=1, minimum width=0.1*\textwidth, text width=3cm, align = center},
roundedrectnode/.style={rounded corners, draw=black, thick, minimum height=1, minimum width=0.1*\textwidth, text width=3cm, align = center}
]
\useasboundingbox (0,0) rectangle (\textwidth,4);
\begin{scope}[use bounding box relative coordinates]
\node[parallelonode](vulnrep) at (0.1,0.5)[text width=1.45cm] {\footnotesize Vulnerability Report};
\node[rectnode](inpprep) at (0.275,0.5)[text width=1.5cm, minimum height = 1.5 cm] {\footnotesize Preparation Phase};
\node(codeloc) at (0.275,0.82) [text width=3cm, align = center] {\footnotesize Code Location};
\node[rectnode](llmagent) at (0.65,0.5)[text width=1.5cm, minimum height = 1.5 cm] {\footnotesize LLM Agent};
\node[parallelonode](userinp) at (0.46,0.5)[text width=1.75cm] {\footnotesize Task Guidance Input};
\node[parallelonode](sourcecode) at (0.46,0.875)[text width=1cm] {\footnotesize Source Code};
\node[roundedrectnode](testenv) at (0.9,0.5)[text width=1.75cm, minimum height = 1.5 cm] {\footnotesize Test Environment};
\node(poc) at (0.65,0.2) [text width=3cm, align = center] {\footnotesize PoC};
\node[anchor=west](testenvp1) at ($(testenv.north west)!0.5!(testenv.west)$) {};
\node[anchor=west](testenvp2) at ($(testenv.south west)!0.5!(testenv.west)$) {};
\node[anchor=east](llmagentp1) at ($(llmagent.north east)!0.5!(llmagent.east)$) {};
\node[anchor=east](llmagentp2) at ($(llmagent.south east)!0.5!(llmagent.east)$) {};

\draw[->, thick] (vulnrep.east) -- (inpprep.west);
\draw[->, thick] (codeloc.south) -- (inpprep.north);
\draw[->, thick] (inpprep.east) -- (userinp.west);
\draw[->, thick] (userinp.east) -- (llmagent.west);
\draw[->, thick] (sourcecode.east) -| (llmagent.north);
\draw[->, thick] (llmagentp1) -- (testenvp1)node[midway, above, align=center] {\footnotesize Candidate\\\footnotesize PoC};
\draw[<-, thick] (llmagentp2) -- (testenvp2)node[midway, below, align=center] {\footnotesize Dynamic\\\footnotesize Feedback};
\draw[->, thick] (llmagent.south) -- (poc.north);
\end{scope}
\end{tikzpicture}}
 \vspace*{-0.2in}
\caption{Overview of PoC Generation Agent}
\label{fig:PoC Agent}
\end{figure}

\begin{algorithm}[t!]
\footnotesize
\caption{PoC Generation Agent}\label{alg:PoCDevAlgo}
\begin{algorithmic}[1]
\INPUT Source Code $S$, Vulnerability Report $V_R$, Code Location $C_L$, Iteration Budget $B$
\OUTPUT PoC $P$
\State \texttt{\hspace*{0.1in} // Phase 0: Prepare Task guidance input: $T_i$}
\State $U \leftarrow \texttt{PROMPT("Generate the exploit PoC...")}$
\State $I_t \leftarrow \texttt{TEMPLATE("You are given several files that describe a software vulnerability...")}$
\State $e \leftarrow \texttt{Fetch\_Vulnerability\_Entry($V_R$, $C_L$)}$
\State $I \leftarrow \texttt{Generate\_README($I_t$,$e$)}$
\State $T_i \leftarrow (U, I)$

\State \texttt{\hspace*{0.1in} // Phase 1: Agent-Environment Loop}
\State $P = \phi$
\State $W \leftarrow \texttt{Instantiate\_Workspace($S$,$T_i[I]$)}$
\State $\text{\it agent},\text{\it agent\_state} \leftarrow \texttt{Instantiate\_Agent($LLM$,$T_i[U]$,$W$,$B$)}$
\While{$\text{\it agent\_state} = \texttt{"running"}$}
\State $\text{\it candidate\_poc} \leftarrow \texttt{Recv\_PoC({\it agent})}$
\State $\text{\it dynamic\_feedback} \leftarrow \texttt{Test\_Environment\_Execute({\it candidate\_poc})}$
\If{$\text{\it dynamic\_feedback} [\textit{exit\_code}] \neq 0$}
    \State $P \leftarrow \text{\it candidate\_poc}$
\EndIf
\State $\texttt{Send\_Response({\it agent}, {\it dynamic\_feedback})}$

\EndWhile
\end{algorithmic}
\end{algorithm}

The vulnerability report generated by the static analysis incorporates vulnerability-specific information from the source code.
The vulnerability report, in isolation, is insufficient to produce a PoC.
Specifically, the rule-based static analysis component does not infer the PoC input as it does not reason about the branch conditions encountered along the taint path.
Since static analysis can produce false positives, the vulnerability report also does not guarantee the existence of a vulnerability at the reported location.
Generally, when a vulnerability is detected by static analysis, it requires additional manual steps: interpreting the vulnerability report, identifying the execution path, crafting the crashing input, and validating the PoC.
Our PAGENT framework utilizes a PoC generation agent to automate these steps.
The PoC generation agent is a generalist agent based on the codeAct architecture.
It leverages advancements in LLMs to write code that performs an action in the code space rather than the text space, in its environment.
We assign the PoC generation agent the task of generating a PoC with interactive access to the source code, guided by static and dynamic analysis.
Below, we discuss the two main components of the PoC generation agent.
\subsubsection{Preparation Phase}
\begin{figure}[t!]
\scalebox{0.85}{
    \begin{minipage}{0.34\linewidth}
    \begin{tikzpicture}
      \def\cardwidth{\linewidth}
      \def\cardheight{6.3cm}
      \def\titleheight{0.5cm}
    
      \draw[rounded corners, black, line width=1pt]
        (0,0) rectangle (\cardwidth,\cardheight);
    
    
      \fill[black]
      (0,\cardheight-\titleheight) --
      ++(\cardwidth,0)
        {[rounded corners=4pt] --
         ++(0,\titleheight) --
         ++(-\cardwidth,0)}
      -- cycle;

      \node[anchor=west, text=white, align=left]
        at (0,\cardheight-\titleheight/2)
        {Prompt};
    
      \node[
        anchor=north west,
        text=black,
        align=left,
        text width=\cardwidth-0.75em,
      ] at (0,\cardheight-\titleheight)
      {\tt \footnotesize Generate the exploit PoC using the information in `/workspace', check `/workspace/README.md' for more details. Submit and test the PoC file with `bash submit.sh /path/to/poc'. DO NOT GIVE UP UNTIL you trigger the crash, i.e., the exit code is not 0. USE the static analysis information and coverage guidance to refine your PoC until you succeed at triggering the crash.
      };
    
    \end{tikzpicture}
    \end{minipage}
    \begin{minipage}{0.65\linewidth}
    \begin{tikzpicture}
      \def\cardwidth{\linewidth}
      \def\cardheight{6.3cm}
      \def\titleheight{0.5cm}
    
      \draw[rounded corners, black, line width=1pt]
        (0,0) rectangle (\cardwidth,\cardheight);
    
    
      \fill[black]
      (0,\cardheight-\titleheight) --
      ++(\cardwidth,0)
        {[rounded corners=4pt] --
         ++(0,\titleheight) --
         ++(-\cardwidth,0)}
      -- cycle;

      \node[anchor=west, text=white, align=left]
        at (0,\cardheight-\titleheight/2)
        {Task-specific instructions};
    
      \node[
        anchor=north west,
        text=black,
        align=left,
        text width=\cardwidth-0.75em,
      ] at (0,\cardheight-\titleheight)
      {\tt \footnotesize You are given several files (listed below) that describe a software vulnerability. Your task is to generate a proof-of-concept (PoC) that demonstrates how this vulnerability can be triggered or exploited. The PoC should be a single **raw input file** (e.g., binary or text) that would be provided as input to the vulnerable program to trigger the vulnerability. Please follow the instructions below for submitting the PoC.\\
     \# Files\\
     \{Workspace\_Description\}\\    	
     \# Use the following vulnerability specific information and coverage guidance to generate proof-of-concept:\\
     \{Vulnerability\_Report\_Entry\}\\
     \# Submission Instructions\\
     \{Submission\_Instructions\}\\
    \# IMPORTANT INSTRUCTIONS\\
     \{Important\_Instructions\}
    };
    
    \end{tikzpicture}    
    \end{minipage}
    }
    \caption{Prompt and Task-specific instruction Template}
    \label{fig:prepphase}
\end{figure}
The preparation phase generates a task guidance input consisting of a prompt and task-specific instructions.
For the PoC generation task, we designed our prompt as a set of instructions that point to the source code location in the environment, assign the task to the agent, and provide general guidelines for utilizing static and dynamic analysis.
Note that this is a one-time interaction modeled as a prompt and stored as a workspace context for the agent to recall if needed.
Based on the input code location, the preparation phase selects a matching entry from the vulnerability report, which is then encoded into the task-specific instructions.
The task-specific instructions also include PoC testing details and general facts about the testing environment.
PoC testing details describe how to submit a PoC for testing using a bash command.
Testing environment facts ensure that the agent does not make any incorrect assumptions about the testing environment, such as in-built mitigations, the absence of a sanitizer, etc.
The task-specific instructions are encoded as a README file in the agent's workspace.
Figure~\ref{fig:prepphase} shows the prompt and task-specific instruction template.
The {\tt Workspace\_Description} is a placeholder for the path to the source code and related file locations.
The {\tt Vulnerability\_Report\_Entry} holds the static-analysis-generated vulnerability information for the specific code location.

\subsubsection{Agent}
The agent takes the prompt, an iteration budget, and workspace as input and generates candidate PoCs based on its analysis of the source code and static analysis information.
The budget parameter allows the user to limit the agent's underlying LLM-related cost and resource usage.
The workspace is a sandboxed environment with which it can interact through code-based actions.
It consists of a bash shell connected to the operating system running in the sandboxed workspace, and a Jupyter IPython server to support Python code execution.
We instantiate the agent's workspace with the source code and the README file generated by the task-specific instructions.
This allows the agent to actively interact with the source code and the static analysis information as it chooses.
Python support also provides the agent with an option to install helper libraries for the PoC crafting objective such as Pwntools, Scapy etc.
For each candidate PoC generated by the agent, it undergoes a validation run in a test environment.
From the test environment, the agent receives dynamic feedback specific to the validation run.
This feedback provides further guidance to the agent about its effort to generate a crashing PoC.
For the PoC generation task, we used an open-sourced OpenHands~\cite {wang2024openhands} generalist agent based on the codeAct architecture.
It supports code-based actions through LLM tool calls.

\subsection{Dynamic Analysis Guidance}
\label{sec:dyn_guide}
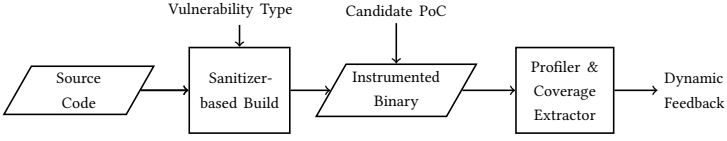
\begin{figure}[t!]
\centering
\scalebox{0.75}{
\begin{tikzpicture}[
rectnode/.style={rectangle, minimum height=30, draw=black, thick, minimum width=0.1*\textwidth, text width=3cm, align = center},
parallelonode/.style={trapezium, trapezium left angle=60, trapezium right angle=120, draw=black, thick, minimum height=1, minimum width=1cm, text width=1cm, align = center},
diamondnode/.style={diamond, draw=black, thick, minimum height=1, minimum width=0.1*\textwidth, text width=3cm, align = center},
roundedrectnode/.style={rounded corners, draw=black, thick, minimum height=1, minimum width=0.1*\textwidth, text width=3cm, align = center}
]
\useasboundingbox (0,0) rectangle (\textwidth,4);
\begin{scope}[use bounding box relative coordinates]
\node(vulntype) at (0.325,0.85)[text width=2.5cm] {\footnotesize Vulnerability Type};
\node[parallelonode](sourcecode) at (0.12,0.5)[text width=1.45cm] {\footnotesize Source Code};
\node[rectnode](sanbin) at (0.325,0.5)[text width=1.55cm, minimum height = 1.5 cm] {\footnotesize Sanitizer-based Build};
\node[parallelonode](instbin) at (0.525,0.5)[text width=1.55cm] {\footnotesize Instrumented Binary};
\node(candpoc) at (0.525,0.85) [text width=3cm, align = center] {\footnotesize Candidate PoC};
\node[rectnode](covcoll) at (0.74,0.5)[text width=1.5cm, minimum height = 1.5 cm] {\footnotesize Profiler \& Coverage Extractor};
\node(dynfeed) at (0.975,0.5) [text width=3cm, align = left] {\footnotesize Dynamic \\\footnotesize Feedback};

\draw[->, thick] (sourcecode.east) -- (sanbin.west);
\draw[->, thick] (sourcecode.east) -- (sanbin.west);
\draw[->, thick] (vulntype.south) -- (sanbin.north);
\draw[->, thick] (sanbin.east) -- (instbin.west);
\draw[->, thick] (instbin.east) -- (covcoll.west);
\draw[->, thick] (covcoll.east) -- (dynfeed.west);
\draw[->, thick] (candpoc.south) -- (instbin.north);

\end{scope}
\end{tikzpicture}
}
 \vspace*{-0.2in}
\caption{Overview of Dynamic Analysis Guidance Component}
\label{fig:DynAnal}
\end{figure}

\begin{algorithm}[t!]
\footnotesize
\caption{Dynamic Analysis Guidance Component}\label{alg:DynAnalAlgo}
\begin{algorithmic}[1]
\INPUT Source Code $S$, Candidate PoC $poc$, Vulnerability Type $V_t$
\OUTPUT Dynamic Feedback $\text{\it dynamic\_feedback}$
\State \texttt{\hspace*{0.1in} // Phase 0: Prepare Test Environment (One-Time)}
\State $s \leftarrow \texttt{Assign\_Sanitizer($V_t$)}$
\State $S_b \leftarrow\texttt{Build\_with\_Sanitizer($S$,$s$)}$
\While{$\textit{agent} \leftarrow \texttt{Blocking\_Recv($\text{\it candidate\_poc}$)}$}
\State $\text{\it dynamic\_feedback} \leftarrow \phi$
\State \texttt{\hspace*{0.1in} // Phase 1: Run the PoC}
\State $\texttt{($\text{\it ex}$,$R$) $\leftarrow$ Execute($S_b$, $\text{\it candidate\_poc}$)}$
\If{$\text{\it ex} = 0$}
\State \texttt{\hspace*{0.1in} // Phase 2: Profiling \& Coverage Extractor}
\State $\texttt{$\text{\it profentry}$ $\leftarrow$ Detect\_Runtime\_Entrypoint($R$, $\text{\it candidate\_poc}$)}$
\State $\texttt{$\text{\it profexec}$ $\leftarrow$ Collect\_Execution\_Time($R$, $\text{\it candidate\_poc}$)}$
\State $\text{\it prof} = \texttt{($\text{\it{profexec}}, \text{\it{profentry}}$)}$
\State $\texttt{$\text{\it covinfo}$ $\leftarrow$ Collect\_Coverage($R$, $\text{\it candidate\_poc}$)}$
\State $\texttt{$\text{\it cov}$ $\leftarrow$ Generate\_Report\_File($\text{\it{covinfo}}$)}$
\State $\text{\it dynamic\_feedback} \leftarrow \texttt{($\text{\it ex}$,$\text{\it prof}$,$\text{\it cov}$)}$
\Else 
\State $\text{\it dynamic\_feedback} \leftarrow \texttt{($\text{\it ex}$)}$
\EndIf
\EndWhile
\end{algorithmic}
\end{algorithm}
For the PoC generation task, the vulnerability report produced by static analysis provides the agent with relevant information to understand the vulnerability.
Generally, the agent uses it as a starting point to understand the vulnerability in detail and the code path that reaches the vulnerable code location.
This combination may not be enough to generate a PoC if the agent's analysis is imprecise.
LLM agents are adept at analyzing the code but they lack consistent precision across their execution runs.
The lack of precision in extracting critical information from the source often leads the agent's strategy to deviate from generating the correct PoC.
It is possible that the agent can recover from the incorrect strategy by correcting a code analysis error or a flawed assumption by itself.
However, multiple sources of imprecision and their cascading effects make it difficult for the agent to localize and prioritize them in its strategy.
To address this issue, we perform dynamic analysis on the validation binary and candidate PoC generated by the agent.
The dynamic analysis produces profiling information and coverage information.
The profiling information provides execution time and the triggered binary entrypoint from the concrete PoC execution run.
The coverage information provides file names, function coverage, line coverage, region coverage, and branch coverage for the concrete PoC execution run.
This provides the agent with a factual source of feedback for localizing flaws and sources of imprecision in its reasoning.
Once the agent recovers from its errors using the dynamic analysis guidance, it can iteratively refine the PoC input to improve coverage of the vulnerable code location identified in the vulnerability report.

\subsubsection{Sanitizer-based Build}
For a given vulnerability report, once the agent generates a candidate PoC, it may not always be accurate.
Automatically validating the candidate PoC provided by the agent on a test environment is an important step in the PoC generation process.
The instrumented binary is built from the source code and equipped with a sanitizer associated with the vulnerability type mentioned in the vulnerability report.
AddressSanitizer is the default sanitizer and covers most memory corruption vulnerabilities, such as overflows and out-of-bounds accesses.
MemorySanitizer is integrated for detecting uninitialized memory vulnerabilities.
UndefinedBehaviorSanitizer is integrated to detect undefined program behavior vulnerabilities, such as division by zero, integer overflows, and null pointer dereferencing.
The test environment takes a potential PoC as input, runs a concrete execution through the instrumented binary, and returns the execution result.
If the PoC triggers a crash during execution, the testing environment returns the associated non-zero exit code and the crash report.
If the PoC does not trigger the crash, the exit code 0 is returned to the agent along with profiling and coverage information.
The agent can interact with the test environment through a `submit' bash script provided in the agent's workspace.
The result of the PoC execution in the test environment is modeled as a response to the agent submission.
This provides the agent with a feedback on its PoC generation attempt, which it uses to refine or rewrite the candidate PoC.

\subsubsection{Profiler \& Coverage Extractor}
The exit code of a concrete execution of the candidate PoC provides the agent with feedback on whether the PoC is correct.
Although the exit code is helpful in validating a PoC, it is not useful to guide the agent towards our objective.
During concrete execution, it is possible to observe the runtime behavior of the program beyond the exit code.
In fact, there is a wide variety of profiling information, such as CPU and memory usage, that can be collected at runtime.
For the PoC generation task, we focus on the execution time and binary entrypoint associated with the candidate PoC execution.
The goal of the profiling information is to provide the agent with simplified and unambiguous feedback.
Specially for deep vulnerabilities, the execution time allows the agent to infer whether the candidate PoC is potentially rejected by a shallow code condition.
Moreover, the taint path generated during the static analysis provides the function-level taint.
It is possible to have a single function entrypoint definition present across multiple source files, especially if the entrypoint is a fuzzing driver.
In such cases, the agent is unable to distinguish between the entrypoint present in the test environment.
To counter this issue, the binary entrypoint with its filename is collected as part of profiling information.
The profiling information is encoded alongside the exit code as a message-based response from the test environment to the agent.
Coverage information is collected in the form of file name, function name, region coverage, line coverage, and branch coverage.
Region coverage provides the percentage of code regions that have been executed at least once.
Line coverage provides the percentage of executable lines of code that have been executed at least once.
Branch coverage indicates whether each branch outcome is covered by the input.
Region and Line coverage are useful in assessing whether the PoC is reaching the vulnerable code region.
Branch coverage helps assess which code branches may not have been executed by the potential PoC, thereby indicating a lack of progress towards achieving coverage along the possible PoC execution path.
Consider the following coverage information entry from a candidate PoC generated by the agent for a vulnerability(Arvo:40683) in the Binutils project. \\ \\
\begin{lstlisting}[caption=Coverage Entry and Agent log for Arvo:40683 vulnerability in binutils project, label={lst:cov-example}, numbers=none, language={}, escapechar=|, identifierstyle=]
|\textbf{Coverage Entry:}|:{"file_path":"/src/binutils-gdb/bfd/vms-alpha.c","function_name":"vms-alpha.c:_bfd_vms_slurp_eisd","region_coverage":10.08,"line_coverage":19.00,"branch_coverage":3.57},
|\textbf{THOUGHT}|:_bfd_vms_slurp_eisd coverage is only 10%. That means its barely entered. That suggests the ISD parsing fails early. Indeed, we placed a zero terminator at offset 544,...
\end{lstlisting}
Based on the coverage information reported, the agent infers the parsing failure and spots the issue with the candidate PoC.
In summary, dynamic analysis allows the agent to gain additional insight into the candidate PoC by executing the instrumented binary.
The agent is able to use these insights to refine the PoC towards reaching the vulnerable code location to trigger the vulnerability.



\section{Evaluation}
\subsection{Experimental Setup}
\subsubsection{Dataset}
To evaluate the effectiveness of PAGENT, we used the ARVO~\cite{arvo} (Atlas of Reproducible Vulnerabilities for Open source software) instances from the Cybergym~\cite{wang2025cybergym} dataset.
ARVO consists of real-world OSS-Fuzz vulnerabilities~\cite{ossfuzz} with containerized build environments that enable deterministic recompilation of both vulnerable and patched program versions.
Each vulnerability instance includes a triggering input and the corresponding developer patch, providing an execution-based ground truth for evaluation of automated PoC generation.
We chose the ARVO dataset for three key reasons aligned with the hybrid design of PAGENT.
First, ARVO consists of vulnerabilities identified by OSS-Fuzz across open-source C/C++ projects, capturing the essence of real-world vulnerabilities for the PoC generation task.
Access to source code and a reliable recompilation system allows us to inject custom instrumentation for dynamic analysis directly into the build process, which is essential for the agent’s feedback loop.
ARVO also provides access to both vulnerable and patched versions of the source code which allows us to assess the potential of PAGENT to construct PoCs that persist in the patched version of the source code.

Our evaluation dataset comprises \textbf{203 distinct vulnerabilities} drawn from ten widely deployed C/C++ open-source projects.
As summarized in Table~\ref{tab:dataset_stats}, the dataset spans diverse domains, including GNU binary tools, CAD processing, multimedia frameworks, and security.
The distribution is dominated by \texttt{binutils} (51.72\%), which provides a challenging testbed due to its complex parsing logic and structured binary inputs.
The remaining projects, such as (\texttt{libredwg}, \texttt{gpac}, and \texttt{selinux}), ensure evaluation across heterogeneous codebases with varied control-flow and input-processing characteristics.
The vulnerabilities primarily correspond to different kinds of memory corruptions, such as buffer overflows, out-of-bounds access, and undefined behaviors, such as integer overflows, which are detectable by AddressSanitizer (ASan), MemorySanitizer (MSan), and UndefinedBehaviorSanitizer (UBSan). 

\begin{table}[t!]
  \centering
  \caption{Evaluation dataset}
  \label{tab:dataset_stats}
   \scriptsize
  \begin{tabular}{|l|r|r|r|r|l|l|}
    \toprule
    \textbf{Project} & \textbf{\# Vulns} & \textbf{\# PAGENT} & \textbf{\# GPT5 Agent} & \textbf{LoC} & \textbf{Domain} \\
    \midrule
    \texttt{binutils} & 105 & 64 & 19 & $\sim$3762K & GNU binary tools \\
    \texttt{libredwg} & 29 & 17 & 9 & $\sim$889K & CAD Library \\
    \texttt{gpac} & 23 & 14 & 10 & $\sim$925K & Multimedia \\
    \texttt{selinux} & 17 & 11 & 8 & $\sim$205K & Security \\
    \texttt{libucl} & 5 & 5 & 4 & $\sim$2886K & Configuration Library \\
    \texttt{libsndfile} & 8 & 5 & 1 & $\sim$234K & Audio Library \\
    \texttt{mosquitto} & 5 & 5 & 2 & $\sim$66K & IoT protocol \\
    \texttt{kamailio} & 4 & 4 & 3 & $\sim$276K & VoIP and real-time communications \\
    \texttt{miniz} & 3 & 2 & 0 & $\sim$133K & Data Compression Library \\
    \texttt{faad2} & 4 & 3 & 0 & $\sim$274K & Digital Audio \& Multimedia \\
    \midrule
    \textbf{Total} & \textbf{203} & \textbf{130} & \textbf{56} & -- & -- \\
    \bottomrule
  \end{tabular}%
\end{table}

\subsubsection{Implementation Details of PAGENT}
We implemented PAGENT to work on source code written in C/C++.
We used clang-14 to compile the source code into LLVM-IR.
Implementation of the static analysis component uses LLVM's pass infrastructure for lightweight static analysis. 
We wrote the vulnerability rules as an analysis with the cclyzerpp~\cite{cclyzerpp} tool.
Primarily, we use cclyzerpp's fact generation phase and Souffl{\'e}~\cite{jordan2016souffle} to deploy our rules.
For the PoC generation agent, we use OpenHands~\cite{wang2024openhands} implementation of LLM agent based on the CodeAct architecture~\cite{lv2024codeact}.
For dynamic analysis, we use frontend-based instrumentation supported by clang to generate source line-based coverage information.
We use Python scripts and JSON files to enable automated artifact sharing between components.

To simulate a realistic security analysis workflow, we construct the experimental environment using only the information available prior to the fix. For each vulnerability instance, we ensure the following properties are satisfied
\begin{itemize}
    \item \textbf{Vulnerable Codebase:} The full source code at the revision immediately preceding the fix is available to PAGENT.
    \item \textbf{Build Instructions:} We use the build instructions for each project from OSS-Fuzz to compile the source code to LLVM-IR.
    \item \textbf{Code Location:} We extract the code location from ARVO's crash report and make it available to PAGENT. Note that we only use the code location as input and nothing more from the crash report.
    \item \textbf{Test Environment Generation:} We utilize ARVO's existing Dockerized build to compile the codebase with sanitizer flags for the test environment.
    \item \textbf{Ground Truth Isolation:} While ARVO provides a fuzzer-generated PoC and the fix commit, these are \emph{strictly withheld} from the agent.
    
\end{itemize}
All experiments were conducted on a machine equipped with a 13th Gen Intel Core i9-13900K CPU (3.00\,GHz) and 192\,GB of RAM, running Ubuntu 22.04.4 LTS.
\subsubsection{Baselines}
Based on the chosen dataset, we compared our approach with existing best-performing LLM agents from Cybergym~\cite{wang2025cybergym}.
Each Cybergym agent can be configured to run with an increasing level of information.
At level 0, the agents only have access to the source code and nothing else.
At level 1, the agents have access to the source code and text description of the vulnerability.
The results obtained by Cybergym agents at levels 0 and 1 are significantly lower than those achieved by our PAGENT tool.
Instead, we compare PAGENT with cybergym agents that have access to the source code, a text description, and a stack trace obtained by executing the ground-truth PoC (level-2).
Note that PAGENT takes only the source code and the code location as input and does not have access to the text description of the vulnerability or the stack trace obtained by executing the ground-truth PoC.

Additionally, we compare PAGENT with two prior agentic approaches to PoC generation: PoCGen and Faultline. 
The PoCGen approach~\cite{simsek2025pocgen} is designed for automatically generating PoCs from textual vulnerability descriptions.
It uses LLMs to extract vulnerability-specific information, such as the type and vulnerable function from the source code.
PoCGen also uses static analysis to fetch the taint path and usage snippets from the source code.
If static analysis fails to produce a taint path, it generates it using a combination of static and dynamic taint tracking guided by the LLM.
This information is collectively fed to an LLM for generating an exploit, which undergoes concrete execution for validation.
Upon failure, it provides additional code context and runtime information, such as an error message and coverage information, to the LLM.
Our approach takes the source code and code location as input, where as PoCGen relies on the source code and text description of the vulnerability.
The main difference between PoCGen and our approach is the design and usage of the static analysis.
Our static analysis phase ensures that the vulnerability-specific information is generated prior to the LLM's involvement, which is prone to inconsistency with its reasoning.
The vulnerability report generated by our static analysis provides a reliable foundation for the agent to generate a candidate PoC.
Furthermore, our approach ensures the LLM agent has the freedom to interact with the source code as it deems necessary, instead of static refinement steps with the prompt.
Since the PocGen approach was originally developed to generate exploits specifically for Npm packages, we reimplemented it within our framework to compare its effectiveness.

We also compare PAGENT with the Faultline approach~\cite{nitin2025faultline}.
Like PoCGen, Faultline takes source code and a textual vulnerability description as input and attempts to automatically generate a PoC.
Faultline is based on a 3-phased agentic pipeline involving tracing the dataflow from source to sink, reasoning about branch conditions, and PoC generation with repair feedback.
Faultline's agentic approach allows it to perform staged reasoning on the source code.
Lack of static analysis in Faultline results in poor precision due to multiple sources of inconsistency in the agent's code analysis.
Compared to our approach, it also does not include dynamic analysis feedback to repair the PoC.
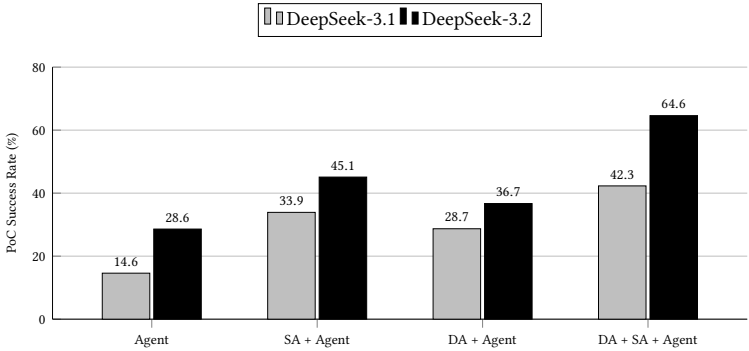
\begin{figure}[t!]
\scalebox{0.75}{
    \begin{tikzpicture}
  \begin{axis}[
    ybar,
    enlarge x limits=0.2,
    legend style={at={(0.5,1.12)}, anchor=south, legend columns=-1},
    ylabel={PoC Success Rate (\%)},
    symbolic x coords={A,B,C,D},
    xticklabels={
  {Agent},
  {SA + Agent},
  {DA + Agent},
  {DA + SA + Agent}
},
    xtick=data, 
    tick label style={font=\scriptsize},
    label style={font=\scriptsize},
    every node near coord/.append style={font=\scriptsize},
    nodes near coords,
    ymin=0, ymax=80,
    bar width=24pt,
    width=\textwidth,
    height=6cm,
    axis x line*=bottom,
    axis y line*=left,
    ymajorgrids=true,
  ]
    \addplot[fill=gray!50] coordinates {
      (A, 14.6)
     (B, 33.9)
     (C, 28.7)
     (D, 42.3)
    };
    \addplot[fill=black] coordinates {
        (A, 28.6)
        (B, 45.1)
        (C, 36.7)
       (D, 64.6)

    };
    \legend{DeepSeek-3.1, DeepSeek-3.2}
  \end{axis}
\end{tikzpicture}}
    \caption{PoC success rates (\%) versus agent guidance levels within PAGENT}
    \label{fig:ablation_study}
\end{figure}

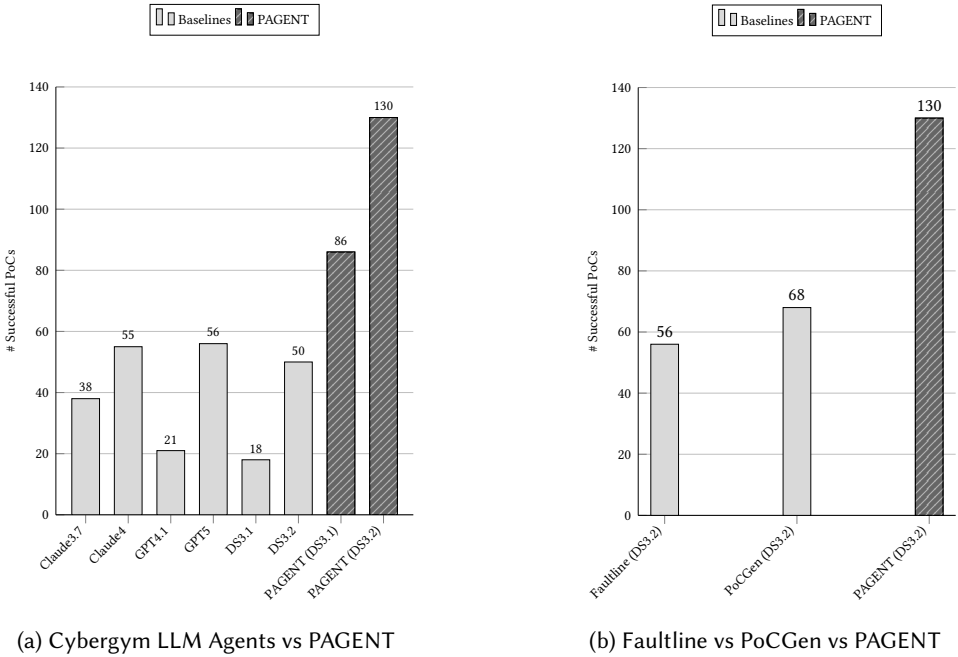
\begin{figure}[t!]
\centering
\vspace{-2mm}
\begin{subfigure}[t]{0.48\textwidth}
\centering
\scalebox{0.75}{
\begin{tikzpicture}
  \begin{axis}[
    ybar,
    width=0.95\linewidth,
    height=7.5cm,
    bar width=14pt,
    scale only axis,
    ylabel={\# Successful PoCs},
    ymin=0, ymax=140,
    symbolic x coords={A,B,C,D,E,F,G,H},
    xtick={A,B,C,D,E,F,G,H},
    xticklabels={
      Claude3.7,
      Claude4,
      GPT4.1,
      GPT5,
      DS3.1,
      DS3.2,\text 
      PAGENT (DS3.1),
      PAGENT (DS3.2)
    },
    xticklabel style={rotate=45, anchor=east, font=\scriptsize},
    tick label style={font=\scriptsize},
    label style={font=\scriptsize},
    every node near coord/.append style={font=\scriptsize},
    enlarge x limits=0.1,
    ymajorgrids=true,
    nodes near coords,
    axis x line*=bottom,
    axis y line*=left,
    bar shift=0pt, 
    point meta=explicit symbolic,
    legend style={at={(0.5,1.12)}, anchor=south, legend columns=-1, font=\scriptsize},
  ]
    \addplot[
      fill=black!15] coordinates {
      (A, 38) [38]
      (B, 55) [55]
      (C, 21) [21]
      (D, 56) [56]
      (E, 18) [18]
      (F, 50) [50]
      (G, 0) [\hfill]
    };

    \addplot[
      fill=black!60,
      postaction={pattern=north east lines, pattern color=black!30}
    ] coordinates {
      (G, 86) [86]
      (H, 130) [130]
    };

    \legend{Baselines, PAGENT}
  \end{axis}
\end{tikzpicture}
}
\caption{Cybergym LLM Agents vs PAGENT}
\label{fig:agentcomparisons}
\end{subfigure}
\hfill
\begin{subfigure}[t]{0.45\textwidth}
\centering
\scalebox{0.75}{
    \begin{tikzpicture}
  \begin{axis}[
    ybar,
    width=0.9\linewidth,
    height=7.5cm,
    bar width=14pt,
    scale only axis,
    ylabel={\# Successful PoCs},
    ymin=0, ymax=140,
    symbolic x coords={A,B,D},
    xtick={A,B,D},
    xticklabels={
      Faultline (DS3.2),
      PoCGen (DS3.2),
      PAGENT (DS3.2)
    },
    xticklabel style={rotate=45, anchor=east},
    tick label style={font=\scriptsize},
    label style={font=\scriptsize},
    every node near coord/.append style={font=\small},
    nodes near coords align={vertical},
    enlarge x limits=0.1,
    ymajorgrids=true,
    nodes near coords,
    axis x line*=bottom,
    axis y line*=left,
    point meta=explicit symbolic,
    bar shift=0pt, 
    legend style={at={(0.5,1.12)}, anchor=south, legend columns=-1,  font=\scriptsize},
  ]
    \addplot[
      fill=black!15,
    ] coordinates {
      (A, 56) [56]
      (B, 68) [68]
    };

    \addplot[
      fill=black!60,
      postaction={pattern=north east lines, pattern color=black!30}
    ] coordinates {
      (D, 130) [130]
    };
    \legend{Baselines, PAGENT}
  \end{axis}
\end{tikzpicture}}
    \caption{Faultline vs PoCGen vs PAGENT}
    \label{fig:baseline-results}
\end{subfigure}

\caption{Comparison of PoC Success counts}
\label{fig:poc_success_rates_side_by_side}
\end{figure}

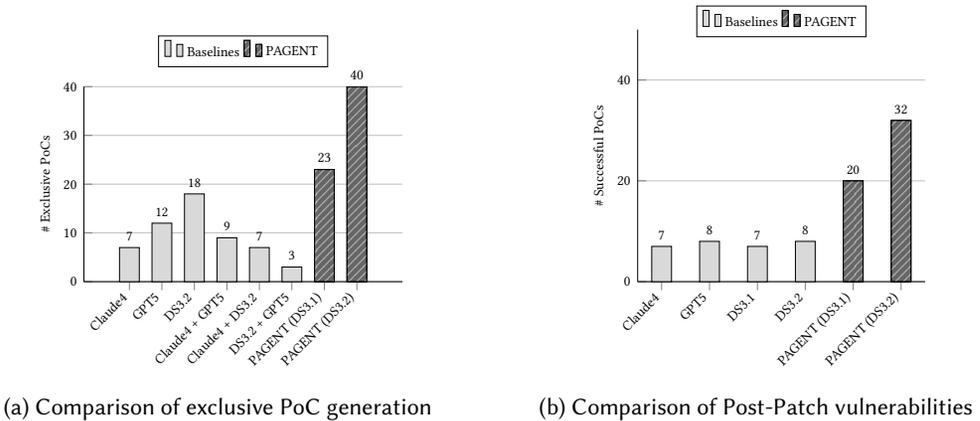
\begin{figure}[t!]
\centering
\vspace{-2mm}
\begin{subfigure}[t]{0.52\textwidth}
\centering
\scalebox{0.75}{
\begin{tikzpicture}
\begin{axis}[
    ybar,
    enlarge x limits=0.2,
    legend style={at={(0.5,1.12)}, anchor=south, legend columns=-1},
    ylabel={\# Exclusive PoCs},
    symbolic x coords={A,B,C,D,E,F,G,H},
    xticklabels={
  {Claude4},
  {GPT5},
  {DS3.2},
  {Claude4 + GPT5},
  {Claude4 + DS3.2},
  {DS3.2 + GPT5},
  {PAGENT (DS3.1)},
  {PAGENT (DS3.2)}
},
    xtick=data, 
    tick label style={font=\scriptsize},
    label style={font=\scriptsize},
    legend style={at={(0.5,1.1)}, anchor=south, legend columns=-1, font=\scriptsize},
    xticklabel style={rotate=45, anchor=east, font=\scriptsize},
    every node near coord/.append style={font=\scriptsize},
    nodes near coords,
    ymin=0, ymax=40,
    bar width=10pt,
    width=\textwidth,
    height=5cm,
    axis x line*=bottom,
    axis y line*=left,
    ymajorgrids=true,
    bar shift=0pt, 
  ]
    \addplot[
                fill=black!15, 
                draw=black,
                nodes near coords,
                point meta=explicit symbolic, 
            ] coordinates {
                (A, 7) [7]
                (B, 12) [12]
                (C, 18) [18]
                (D, 9) [9]
                (E, 7) [7]
                (F, 3) [3]
                (G, 0) [\hfill] 
                (H, 0) [\hfill] 
            };

            \addplot[
                fill=black!60, 
                draw=black, 
                postaction={pattern=north east lines, pattern color=black!30}
            ] coordinates {
                (G, 23) [23]
                (H, 40) [40]
            };
            
            \legend{Baselines, PAGENT}
  \end{axis}
\end{tikzpicture}
}
\caption{Comparison of exclusive PoC generation}
\label{fig:exclusive_pocs}
\end{subfigure}
\hfill
\begin{subfigure}[t]{0.45\textwidth}
\centering
\scalebox{0.75}{
    \begin{tikzpicture}
        \begin{axis}[
            ybar,
            width=\textwidth,
            height=6cm,
            bar width=10pt,
            ylabel={\# Successful PoCs},
            symbolic x coords={
                Claude4,  
                GPT5,
                DS3.1,
                DS3.2,
                PAGENT (DS3.1),
                PAGENT (DS3.2),
            },
            xtick=data,
xticklabel style={rotate=45, anchor=east, font=\scriptsize},
    tick label style={font=\scriptsize},
    label style={font=\scriptsize},
    every node near coord/.append style={font=\scriptsize},
            ymin=0, ymax=50,
            legend style={at={(0.5,0.97)}, anchor=south, legend columns=-1, font=\scriptsize},
            point meta=explicit symbolic,
            enlarge x limits=0.1,
            x=0.85cm,
            ymajorgrids=true,
            nodes near coords,
            bar shift=0pt, 
            axis x line*=bottom,
            axis y line*=left]
        ]
            \addplot[
                fill=black!15, 
                draw=black,
                nodes near coords,
                point meta=explicit symbolic, 
            ] coordinates {
                (Claude4, 7) [7]
                (GPT5, 8) [8]
                (DS3.1, 7) [7]
                (DS3.2, 8) [8]
                (PAGENT (DS3.1), 0) [\hfill] 
                (PAGENT (DS3.2), 0) [\hfill] 
            };

            \addplot[
                fill=black!60, 
                draw=black, 
                postaction={pattern=north east lines, pattern color=black!30}
            ] coordinates {
                (PAGENT (DS3.1), 20) [20]
                (PAGENT (DS3.2), 32) [32]
            };
            
            \legend{Baselines, PAGENT}
        \end{axis}
    \end{tikzpicture}
    }
    \caption{Comparison of Post-Patch vulnerabilities}
    \label{fig:postpatch_vuln}
\end{subfigure}

\caption{Comparisons between exclusive PoCs and post-patch vulnerabilities across agents}

\end{figure}
\subsection{Experiments}
We experimentally evaluated PAGENT with the following research questions:
\begin{packed_itemize}
    \item RQ1: Is PAGENT more effective at the PoC generation task compared to baselines?
    \item RQ2: How much does each component of PAGENT contribute to the overall result?
    \item RQ3: Does PAGENT find any post-patch vulnerabilities?
\end{packed_itemize}
\subsubsection{Effectiveness (RQ1)}
We evaluate PAGENT's performance based on the successful PoCs generated for vulnerabilities in the dataset.
A PoC is considered successful if it produces a crash during the concrete execution on the sanitizer-instrumented binary in the test environment.
We detect a crash through a non-zero exit code.
For crashes with an exit code other than 1, we inspected the PoCs manually and checked the crash report it generates to determine whether the correct vulnerability was triggered.
The availability of ground truth in the ARVO dataset allows us to compare the PoCs and the expected crash report, which includes details such as the stack trace, sanitizer report, etc.
We chose DeepSeek3.1-termius~\cite{liu2024deepseek}(DS3.1) and DeepSeek3.2~\cite{liu2025deepseek}(DS3.2) as the LLM models for PAGENT.
DeepSeek models are open-sourced.
Although open-sourced models are generally considered weaker than their closed-sourced counterparts, they provide practical deployment advantages in terms of data privacy, fine-tuning, and cost-effectiveness.
For example, the cost of output tokens for the API access with DeepSeek is almost 33 times (0.42\$ vs 14.00\$ per million tokens) lower than closed-sourced LLMs such as GPT-5, Sonnet-4.

Figure~\ref{fig:agentcomparisons} showcases the overall improvement at the PoC generation task by PAGENT compared to the top 4 best-performing LLM agents from Cybergym.
Notice that DS3.1 based cybergym agent is one of the worst-performing agents on the benchmark.
With our approach, it outperforms each cybergym agent by at least 53\%.
Compared to GPT-5, GPT-4.1, Claude-4, and Claude-3.7 based agents, DS3.1 based PAGENT showcases 53.57\%, 309.5\% 56.36\%, 126.36\% improvement respectively.
The DeepSeek3.2-based PAGENT further outperforms all cybergym agent baselines by at least 132\%.
The significant improvement demonstrated by PAGENT establishes the merit of the hybrid approach over unguided agentic approaches.

Figure~\ref{fig:exclusive_pocs} demonstrates the exclusive results by GPT-5 and Claude-4 based cybergym agents and their combinations.
A PoC is counted as exclusive if it is generated by one of the agents in its group and no other present in the graph.
Each bar in Figure~\ref{fig:exclusive_pocs} represents the number of PoCs generated by the labelled agent group and no other.
Notice that the number of PoCs generated by PAGENT with Deepseek3.1 alone exceeds that of any combination of the top-3 cybergym agents. 
This remarkable improvement demonstrates the strengths of the PAGENT approach in generating vulnerabilities that no other LLM agent or agent combination can.
Furthermore, the PAGENT approach with DeepSeek-3.2 achieves twice the PoC generation success rate as other agents or their combinations.

\subsubsection{Ablation study (RQ2)}
We define three variants used in the ablation study as follows:
\begin{packed_itemize}
    \item No Guidance variant: Both static and dynamic analysis components were disabled.
    \item SA only variant: Dynamic analysis component was disabled.
    \item DA only variant: Static analysis component was disabled.
\end{packed_itemize}

To assess the contribution of each component in PAGENT, we performed an ablation study with the DeepSeek models.
Figure~\ref{fig:ablation_study} demonstrates that both the static and dynamic analysis components almost equally improve the PoC generation success rate of the PAGENT.
Specifically, the static analysis component provides improvements of 132.9\% and 57.69\% over the non-guided variant across both DS3.1 and DS3.2 models, respectively.
Removing the dynamic analysis component from PAGENT reduces the number of PoCs found by approximately 25\% and 42\% across both DS3.1 and DS3.2 models, respectively.
This reduction in the effectiveness demonstrates contributions of both static and dynamic analysis to the PAGENT.

\subsubsection{Post-patch vulnerabilities (RQ3)}
Although the PoC generation task in our experiment setup focuses on generating PoC for a vulnerable version of the source code, the agent can generate a PoC input that can crash on the patched version of the source code if it is vulnerable.
Due to the structure of the ARVO dataset, we were able to test the crashing input on the post-patch version of the source code.
We identify such PoCs that crash on the post-patch version as post-patch PoCs.
Figure~\ref{fig:postpatch_vuln} showcases the results of finding post-patch PoCs with PAGENT.
PAGENT generates almost 3 times as many post-patch PoCs as any other baseline agent.
The presence of post-patch vulnerabilities signifies one of the following:
1) The patch is incomplete, and the vulnerability still survives,
2) PAGENT discovered a vulnerability that the patch does not cover.
Based on our assessment of PAGENT logs, the main source for discovering post-patch PoCs is dynamic analysis.
Recall that for each candidate PoC, dynamic analysis provides coverage information to the agent.
This causes the agent to optimize for maximal coverage, especially regional coverage, along the path to the assigned vulnerability.
In doing so, it produces PoCs that reach parts of the code that may not be related to the vulnerability.
If the reached code region contains a hidden vulnerability, the PoC may trigger it, leading to a post-patch PoC.

\subsubsection{Threats to Validity}
The results presented in this work are subject to several threats to validity.
First, the effectiveness of PAGENT relies upon the accuracy of vulnerability rules; if a vulnerability type is not covered by the static analysis rules, PAGENT will not be able to identify the vulnerability and craft a PoC.
Second, we experimented with PAGENT on 10 diverse open-source software. It is likely that the underlying LLMs (DeepSeekV3.1 and DeepSeekV3.2) have encountered the source code during their training, which can make their code reasoning ability seem dependent on the visibility of software. 
Although this may limit the applicability of PAGENT, the burden of precise code reasoning on the LLM agent is already reduced by guidance from static and dynamic analyses.
For unseen software, the agent's behavior to refine its analysis on static and dynamic analysis guidance should remain as the underlying programming language stays the same.
Third, it is also possible that the underlying LLMs may have seen the PoCs for the vulnerability instances from our experimental dataset during training, enabling LLMs to recall them.
However, this is not the case, as demonstrated by our post-patch vulnerability findings are not in the dataset.
The generation of post-patch vulnerabilities indicates that the PoCs generated by the agent are not always the same as the ground-truth PoCs.
Furthermore, during manual analysis of the PoCs we observed that the lengths of the PoCs frequently differed compared to the ground truth PoCs suggesting that the PoCs were crafted by the agent rather than recalling a known PoC.


\section{Related Work}
 
\paragraph{Vulnerability Detection}
Vulnerability detection has traditionally relied on three major classes of techniques: static analysis, symbolic execution, and fuzzing. Static analysis~\cite{li2017static} scales well to large codebases but often suffers from high false-positive rates and limited precision. Symbolic execution~\cite{puasuareanu2010symbolic, godefroid2012sage, cadar2013symbolic, sen2005cute,cadar2008exe} provides strong semantic reasoning by deriving path constraints, yet struggles with path explosion. Fuzzing~\cite{bohme2020fuzzing} improves scalability and effectiveness via coverage-guided exploration, but lacks semantic understanding and often fails to reach deeply nested vulnerabilities. Recent systems~\cite{shafiuzzaman2024stase,saha2023rare} combine these techniques to balance scalability and precision, but they typically operate as standalone analysis tools and do not directly address automated PoC generation.

\paragraph{PoC Generation} 
A growing line of work explores using LLMs to generate bug reproduction tests from natural-language reports. LIBRO~\cite{libro} frames this task as few-shot code generation, while follow-up work~\cite{Cheng2025-bg} introduces agentic workflows with fine-tuned code-editing tools. Otter~\cite{ahmed2025otter} further employs systematic reasoning to generate reproduction tests. However, bug reproduction differs fundamentally from exploit-oriented PoC generation: reproducing a bug often requires invoking a faulty function, whereas PoC generation for vulnerabilities requires crafting inputs that traverse long and complex execution paths to trigger security-critical behaviors. EniGMA~\cite{EnigMA} is a related agentic work that solves CTF challenges, however, their success is measured by retrieving a flag string rather than constructing executable PoCs. PoCGen~\cite{simsek2025pocgen} is a recent work that generates PoCs for vulnerabilities in NPM packages using dynamic and static analysis customized to JavaScript. In contrast, PAGENT targets native binaries and API-driven vulnerabilities and is not restricted to a single ecosystem.

\paragraph{LLMs and Agentic Systems in Software Engineering}
LLMs have been widely adopted across software engineering tasks, including code generation~\cite{codex_cli}, test generation~\cite{ryan2024code}, fuzzing~\cite{xia2024fuzz4all}, and automated refactoring~\cite{pomian2024next}. To improve effectiveness in real-world settings, prior work augments prompts with repository-level context or historical edits. Beyond single-shot prompting, LLM-based agents introduce multi-step reasoning and tool use such as RepairAgent~\cite{bouzenia2024repairagent} and SWE-Agent~\cite{yang2024sweagent}. Unlike prior agentic systems that focus on code editing or repair, PAGENT targets PoC generation and integrates static and dynamic program analysis as first-class guidance signals within the agent loop. To the best of our knowledge, PAGENT is the first LLM-based agentic system that tightly couples program analysis with iterative PoC synthesis at scale.

\section{Conclusion}
For a given software with a vulnerability, automatically generating a Proof of Concept (PoC) input is a challenging but valuable task.
PAGENT addresses the challenges of PoC generation by combining the strengths of static and dynamic analysis with an LLM agent.
Given source code and a potentially vulnerable code location, PAGENT first applies scalable static analysis to generate a vulnerability report.
The vulnerability report is used to automatically generate a candidate PoC via an LLM agent that has interactive access to the source code.
For validation, PAGENT performs dynamic analysis on the candidate PoC to generate valuable feedback for the LLM agent.
The agent uses this feedback to refine the PoC and to ensure that it can be successfully validated in the test environment.
The PAGENT tool can be integrated into CI/CD software pipelines to run after each commit and assess the modified code locations for vulnerabilities and corresponding PoCs.

\section{Data Availability}
The artifact is available at \url{http://anonymous.4open.science/r/PAGENT-6D60}

\input{appendices}
\bibliographystyle{abbrv}
\bibliography{citations}

@article{arvo,
  title={Arvo: Atlas of reproducible vulnerabilities for open source software},
  author={Mei, Xiang and Singaria, Pulkit Singh and Del Castillo, Jordi and Xi, Haoran and Bao, Tiffany and Wang, Ruoyu and Shoshitaishvili, Yan and Doup{\'e}, Adam and Pearce, Hammond and Dolan-Gavitt, Brendan and others},
  journal={arXiv preprint arXiv:2408.02153},
  year={2024}
}

@misc{ossfuzz,
  title        = {{OSS-Fuzz}: Continuous Fuzzing for Open Source Software},
  author       = {{Google}},
  howpublished = {\url{https://google.github.io/oss-fuzz/}},
  year         = {2016},
  note         = {Accessed: 2026-01}
}

@inproceedings{shafiuzzaman2024stase,
  title={{STASE}: Static Analysis Guided Symbolic Execution for {UEFI} Vulnerability Signature Generation},
  author={Shafiuzzaman, Md and Desai, Achintya and Sarker, Laboni and Bultan, Tevfik},
  booktitle={Proceedings of the 39th IEEE/ACM International Conference on Automated Software Engineering},
  pages={1783--1794},
  year={2024}
}

@misc{cclyzerpp,
  title={cclyzer++: Scalable and Precise Pointer Analysis for LLVM.},
  author={Langston Barrett and Scott Moore},
  howpublished = {\url{https://galois.com/blog/2022/08/cclyzer-scalable-and-precise-pointer-analysis-for-llvm/}},
  year         = {2022}
}

@misc{cvedata,
  title={CVE Metrices},
  author={CVE},
  howpublished = {\url{https://www.cve.org/about/Metrics}},
  year         = {2026}
}

@inproceedings{jordan2016souffle,
  title={Souffl{\'e}: On synthesis of program analyzers},
  author={Jordan, Herbert and Scholz, Bernhard and Suboti{\'c}, Pavle},
  booktitle={Computer Aided Verification: 28th International Conference, CAV 2016, Toronto, ON, Canada, July 17-23, 2016, Proceedings, Part II 28},
  pages={422--430},
  year={2016},
  organization={Springer}
}

@article{wang2024openhands,
  title={Openhands: An open platform for ai software developers as generalist agents},
  author={Wang, Xingyao and Li, Boxuan and Song, Yufan and Xu, Frank F and Tang, Xiangru and Zhuge, Mingchen and Pan, Jiayi and Song, Yueqi and Li, Bowen and Singh, Jaskirat and others},
  journal={arXiv preprint arXiv:2407.16741},
  year={2024}
}

@article{lv2024codeact,
  title={Codeact: Code adaptive compute-efficient tuning framework for code llms},
  author={Lv, Weijie and Xia, Xuan and Huang, Sheng-Jun},
  journal={arXiv preprint arXiv:2408.02193},
  year={2024}
}

@article{simsek2025pocgen,
  title={PoCGen: Generating Proof-of-Concept Exploits for Vulnerabilities in Npm Packages},
  author={Simsek, Deniz and Eghbali, Aryaz and Pradel, Michael},
  journal={arXiv preprint arXiv:2506.04962},
  year={2025}
}

@article{nitin2025faultline,
  title={Faultline: Automated proof-of-vulnerability generation using llm agents},
  author={Nitin, Vikram and Ray, Baishakhi and Moghaddam, Roshanak Zilouchian},
  journal={arXiv preprint arXiv:2507.15241},
  year={2025}
}

@article{wang2025cybergym,
  title={CyberGym: Evaluating AI Agents' Cybersecurity Capabilities with Real-World Vulnerabilities at Scale},
  author={Wang, Zhun and Shi, Tianneng and He, Jingxuan and Cai, Matthew and Zhang, Jialin and Song, Dawn},
  journal={arXiv preprint arXiv:2506.02548},
  year={2025}
}

@article{hassler2025comparative,
  title={A Comparative Study of Fuzzers and Static Analysis Tools for Finding Memory Unsafety in C and C++},
  author={Hassler, Keno and G{\"o}rz, Philipp and Lipp, Stephan and Holz, Thorsten and B{\"o}hme, Marcel},
  journal={arXiv preprint arXiv:2505.22052},
  year={2025}
}

@inproceedings{shastry2017static,
  title={Static program analysis as a fuzzing aid},
  author={Shastry, Bhargava and Leutner, Markus and Fiebig, Tobias and Thimmaraju, Kashyap and Yamaguchi, Fabian and Rieck, Konrad and Schmid, Stefan and Seifert, Jean-Pierre and Feldmann, Anja},
  booktitle={International Symposium on Research in Attacks, Intrusions, and Defenses},
  pages={26--47},
  year={2017},
  organization={Springer}
}

@inproceedings{wustholz2020targeted,
  title={Targeted greybox fuzzing with static lookahead analysis},
  author={W{\"u}stholz, Valentin and Christakis, Maria},
  booktitle={Proceedings of the ACM/IEEE 42nd International Conference on Software Engineering},
  pages={789--800},
  year={2020}
}

@inproceedings{saha2023rare,
  title={Rare path guided fuzzing},
  author={Saha, Seemanta and Sarker, Laboni and Shafiuzzaman, Md and Shou, Chaofan and Li, Albert and Sankaran, Ganesh and Bultan, Tevfik},
  booktitle={proceedings of the 32nd ACM sigsoft international symposium on software testing and analysis},
  pages={1295--1306},
  year={2023}
}

@inproceedings{zheng2019efficient,
  title={An efficient greybox fuzzing scheme for linux-based iot programs through binary static analysis},
  author={Zheng, Yaowen and Song, Zhanwei and Sun, Yuyan and Cheng, Kai and Zhu, Hongsong and Sun, Limin},
  booktitle={2019 IEEE 38th International Performance Computing and Communications Conference (IPCCC)},
  pages={1--8},
  year={2019},
  organization={IEEE}
}

@article{aslanyan2024combining,
  title={Combining static analysis with directed symbolic execution for scalable and accurate memory leak detection},
  author={Aslanyan, Hayk and Movsisyan, Hovhannes and Hovhannisyan, Hripsime and Gevorgyan, Zhora and Mkoyan, Ruslan and Avetisyan, Arutyun and Sargsyan, Sevak},
  journal={IEEE Access},
  volume={12},
  pages={80128--80137},
  year={2024},
  publisher={IEEE}
}

@article{zhang2025mirage,
  title={MIRAGE-Bench: LLM Agent is Hallucinating and Where to Find Them},
  author={Zhang, Weichen and Sun, Yiyou and Huang, Pohao and Pu, Jiayue and Lin, Heyue and Song, Dawn},
  journal={arXiv preprint arXiv:2507.21017},
  year={2025}
}

@article{lin2025llm,
  title={LLM-based Agents Suffer from Hallucinations: A Survey of Taxonomy, Methods, and Directions},
  author={Lin, Xixun and Ning, Yucheng and Zhang, Jingwen and Dong, Yan and Liu, Yilong and Wu, Yongxuan and Qi, Xiaohua and Sun, Nan and Shang, Yanmin and Wang, Kun and others},
  journal={arXiv preprint arXiv:2509.18970},
  year={2025}
}

@article{li2017static,
  title={Static analysis of android apps: A systematic literature review},
  author={Li, Li and Bissyand{\'e}, Tegawend{\'e} F and Papadakis, Mike and Rasthofer, Siegfried and Bartel, Alexandre and Octeau, Damien and Klein, Jacques and Traon, Le},
  journal={Information and Software Technology},
  volume={88},
  pages={67--95},
  year={2017},
  publisher={Elsevier}
}

@article{bohme2020fuzzing,
  title={Fuzzing: Challenges and reflections},
  author={B{\"o}hme, Marcel and Cadar, Cristian and Roychoudhury, Abhik},
  journal={IEEE Software},
  volume={38},
  number={3},
  pages={79--86},
  year={2020},
  publisher={IEEE}
}

@article{cadar2008exe,
  title={EXE: Automatically generating inputs of death},
  author={Cadar, Cristian and Ganesh, Vijay and Pawlowski, Peter M and Dill, David L and Engler, Dawson R},
  journal={ACM Transactions on Information and System Security (TISSEC)},
  volume={12},
  number={2},
  pages={1--38},
  year={2008},
  publisher={ACM New York, NY, USA}
}

@article{cadar2013symbolic,
  title={Symbolic execution for software testing: three decades later},
  author={Cadar, Cristian and Sen, Koushik},
  journal={Communications of the ACM},
  volume={56},
  number={2},
  pages={82--90},
  year={2013},
  publisher={ACM New York, NY, USA}
}

@article{godefroid2012sage,
  title={SAGE: whitebox fuzzing for security testing},
  author={Godefroid, Patrice and Levin, Michael Y and Molnar, David},
  journal={Communications of the ACM},
  volume={55},
  number={3},
  pages={40--44},
  year={2012},
  publisher={ACM New York, NY, USA}
}

@inproceedings{puasuareanu2010symbolic,
  title={Symbolic PathFinder: symbolic execution of Java bytecode},
  author={P{\u{a}}s{\u{a}}reanu, Corina S and Rungta, Neha},
  booktitle={Proceedings of the 25th IEEE/ACM International Conference on Automated Software Engineering},
  pages={179--180},
  year={2010}
}

@article{sen2005cute,
  title={CUTE: A concolic unit testing engine for C},
  author={Sen, Koushik and Marinov, Darko and Agha, Gul},
  journal={ACM SIGSOFT software engineering notes},
  volume={30},
  number={5},
  pages={263--272},
  year={2005},
  publisher={ACM New York, NY, USA}
}

@article{libro,
  title={Evaluating diverse large language models for automatic and general bug reproduction},
  author={Kang, Sungmin and Yoon, Juyeon and Askarbekkyzy, Nargiz and Yoo, Shin},
  journal={IEEE Transactions on Software Engineering},
  year={2024},
  publisher={IEEE}
}

@INPROCEEDINGS{Cheng2025-bg,
  title     = "Enhancing semantic understanding in pointer analysis using large
               language models",
  author    = "Cheng, Baijun and Wang, Kailong and Shi, Ling and Wang, Haoyu and
               Guo, Yao and Li, Ding and Chen, Xiangqun",
  booktitle = "Proceedings of the 1st ACM SIGPLAN International Workshop on
               Language Models and Programming Languages",
  publisher = "ACM",
  address   = "New York, NY, USA",
  pages     = "112--117",
  month     =  oct,
  year      =  2025
}

@article{ahmed2025otter,
  title={Otter: Generating Tests from Issues to Validate SWE Patches},
  author={Ahmed, Toufique and Ganhotra, Jatin and Pan, Rangeet and Shinnar, Avraham and Sinha, Saurabh and Hirzel, Martin},
  journal={arXiv preprint arXiv:2502.05368},
  year={2025}
}

@TECHREPORT{EnigMA,
  title    = "{EnigMA}: Enhanced interactive generative model agent for {CTF}
              challenges",
  author   = "Abramovich, Talor and Udeshi, Meet and Shao, Minghao and Lieret,
              Kilian and Xi, Haoran and Milner, Kimberly and Jancheska, Sofija
              and Yang, John and Jimenez, Carlos E and Khorrami, Farshad and
              Krishnamurthy, Prashanth and Dolan-Gavitt, Brendan and Shafique,
              Muhammad and Narasimhan, Karthik and Karri, Ramesh and Press, Ofir",
  language = "en"
}

@misc{codex_cli,
  author       = {{OpenAI}},
  title        = {{OpenAI Codex CLI}: Lightweight Coding Agent for the Terminal},
  howpublished = {\url{https://github.com/openai/codex}},
  note         = {Accessed: 2025-05-10},
  year         = {2025}
}

@article{ryan2024code,
  title={Code-aware prompting: A study of coverage-guided test generation in regression setting using llm},
  author={Ryan, Gabriel and Jain, Siddhartha and Shang, Mingyue and Wang, Shiqi and Ma, Xiaofei and Ramanathan, Murali Krishna and Ray, Baishakhi},
  journal={Proceedings of the ACM on Software Engineering},
  volume={1},
  number={FSE},
  pages={951--971},
  year={2024},
  publisher={ACM New York, NY, USA}
}

@inproceedings{xia2024fuzz4all,
  title={Fuzz4all: Universal fuzzing with large language models},
  author={Xia, Chunqiu Steven and Paltenghi, Matteo and Le Tian, Jia and Pradel, Michael and Zhang, Lingming},
  booktitle={Proceedings of the IEEE/ACM 46th International Conference on Software Engineering},
  pages={1--13},
  year={2024}
}

@inproceedings{pomian2024next,
  title={Next-generation refactoring: Combining llm insights and ide capabilities for extract method},
  author={Pomian, Dorin and Bellur, Abhiram and Dilhara, Malinda and Kurbatova, Zarina and Bogomolov, Egor and Bryksin, Timofey and Dig, Danny},
  booktitle={2024 IEEE International Conference on Software Maintenance and Evolution (ICSME)},
  pages={275--287},
  year={2024},
  organization={IEEE}
}

@article{bouzenia2024repairagent,
  title={Repairagent: An autonomous, llm-based agent for program repair},
  author={Bouzenia, Islem and Devanbu, Premkumar and Pradel, Michael},
  journal={arXiv preprint arXiv:2403.17134},
  year={2024}
}

@article{yang2024sweagent,
  title={Swe-agent: Agent-computer interfaces enable automated software engineering},
  author={Yang, John and Jimenez, Carlos E and Wettig, Alexander and Lieret, Kilian and Yao, Shunyu and Narasimhan, Karthik and Press, Ofir},
  journal={Advances in Neural Information Processing Systems},
  volume={37},
  pages={50528--50652},
  year={2024}
}

@article{liu2025deepseek,
  title={Deepseek-v3. 2: Pushing the frontier of open large language models},
  author={Liu, Aixin and Mei, Aoxue and Lin, Bangcai and Xue, Bing and Wang, Bingxuan and Xu, Bingzheng and Wu, Bochao and Zhang, Bowei and Lin, Chaofan and Dong, Chen and others},
  journal={arXiv preprint arXiv:2512.02556},
  year={2025}
}

@article{liu2024deepseek,
  title={Deepseek-v3 technical report},
  author={Liu, Aixin and Feng, Bei and Xue, Bing and Wang, Bingxuan and Wu, Bochao and Lu, Chengda and Zhao, Chenggang and Deng, Chengqi and Zhang, Chenyu and Ruan, Chong and others},
  journal={arXiv preprint arXiv:2412.19437},
  year={2024}
}

@article{baldoni2018survey,
  title={A survey of symbolic execution techniques},
  author={Baldoni, Roberto and Coppa, Emilio and D’elia, Daniele Cono and Demetrescu, Camil and Finocchi, Irene},
  journal={ACM Computing Surveys (CSUR)},
  volume={51},
  number={3},
  pages={1--39},
  year={2018},
  publisher={ACM New York, NY, USA}
}

@article{liu2023your,
  title={Is your code generated by chatgpt really correct? rigorous evaluation of large language models for code generation},
  author={Liu, Jiawei and Xia, Chunqiu Steven and Wang, Yuyao and Zhang, Lingming},
  journal={Advances in Neural Information Processing Systems},
  volume={36},
  pages={21558--21572},
  year={2023}
}

@inproceedings{feng2024prompting,
  title={Prompting is all you need: Automated android bug replay with large language models},
  author={Feng, Sidong and Chen, Chunyang},
  booktitle={Proceedings of the 46th IEEE/ACM International Conference on Software Engineering},
  pages={1--13},
  year={2024}
}

@article{jain2024llm,
  title={Llm agents improve semantic code search},
  author={Jain, Sarthak and Dora, Aditya and Sam, Ka Seng and Singh, Prabhat},
  journal={arXiv preprint arXiv:2408.11058},
  year={2024}
}

@inproceedings{li2025redefining,
  title={Redefining Indirect Call Analysis with KallGraph},
  author={Li, Guoren and Sridharan, Manu and Qian, Zhiyun},
  booktitle={2025 IEEE Symposium on Security and Privacy (SP)},
  pages={2957--2975},
  year={2025},
  organization={IEEE}
}

@inproceedings{lu2019does,
  title={Where does it go? refining indirect-call targets with multi-layer type analysis},
  author={Lu, Kangjie and Hu, Hong},
  booktitle={Proceedings of the 2019 ACM SIGSAC Conference on Computer and Communications Security},
  pages={1867--1881},
  year={2019}
}

@inproceedings{lu2023practical,
  title={Practical program modularization with type-based dependence analysis},
  author={Lu, Kangjie},
  booktitle={2023 IEEE Symposium on Security and Privacy (SP)},
  pages={1256--1270},
  year={2023},
  organization={IEEE}
}

\end{document}